\documentclass[10pt,twocolumn,english,aps,prd,superscriptaddress,nofootinbib,preprintnumbers,floatfix]{revtex4-2}

\usepackage[utf8]{inputenc}
\usepackage[T1]{fontenc}
\usepackage[english]{babel}
\usepackage{amsmath,amssymb,amsfonts,bm,mathrsfs}
\usepackage{graphicx}
\usepackage{xcolor}
\colorlet{BLUE}{blue} 
\usepackage{hyperref}
\usepackage{tikz}

\hypersetup{colorlinks=true,linkcolor=blue,citecolor=blue,urlcolor=blue}

\definecolor{lime}{HTML}{A6CE39}
\newcommand{\orcidicon}{%
    \begin{tikzpicture}
    \draw[lime,fill=lime] (0,0) circle [radius=0.16]
        node[white] {{\fontfamily{qag}\selectfont \tiny ID}};
    \draw[white,fill=white] (-0.0625,0.095) circle [radius=0.007];
    \end{tikzpicture}\hspace{-2mm}
}
\newcommand\orcidFrancisco{{\href{https://orcid.org/0000-0002-9388-8373}{\orcidicon}}}
\newcommand\orcidManuel{{\href{https://orcid.org/0000-0001-8586-0285}{\orcidicon}}}

\begin{document}
\title{Stochastic Thermodynamics of Dynamical Thin-Shell Wormholes: Effective Temperatures, Fluctuations and Conditional Quasi-Horizon Unification}

\author{Francisco S. N. Lobo\orcidFrancisco\!\!} 
\email{fslobo@ciencias.ulisboa.pt}
\affiliation{Instituto de Astrof\'{i}sica e Ci\^{e}ncias do Espa\c{c}o, Faculdade de Ci\^{e}ncias da Universidade de Lisboa, Edifício C8, Campo Grande, P-1749-016 Lisbon, Portugal}
\affiliation{Departamento de F\'{i}sica, Faculdade de Ci\^{e}ncias da Universidade de Lisboa, Edif\'{i}cio C8, Campo Grande, P-1749-016 Lisbon, Portugal}

\author{Manuel E. Rodrigues\orcidManuel\!\!}
\email{esialg@gmail.com}
\affiliation{Faculdade de F\'{\i}sica, Programa de P\'{o}s-Gradua\c{c}\~{a}o em F\'isica, Universidade Federal do Par\'{a}, 66075-110, Bel\'{e}m, Par\'{a}, Brazil}
\affiliation{Faculdade de Ci\^{e}ncias Exatas e Tecnologia, Universidade Federal do Par\'{a}, Campus Universit\'{a}rio de Abaetetuba, 68440-000, Abaetetuba, Par\'{a}, Brazil}

\date{\today}

\begin{abstract}

We develop an effective stochastic and semiclassical framework for dynamical thin-shell wormholes joining two asymptotic spacetime sectors. The Israel junction geometry is kept distinct from the phenomenological open-system input: the fundamental stochastic dynamics is formulated in the radial phase space $(a,\dot a)$, while $\sigma$, $P$, and the effective temperature scales are induced observables, and a one-dimensional Langevin--Fokker--Planck process is introduced only after an explicit overdamped/adiabatic closure. The local acceleration scales are used as reservoir temperatures only under additional local KMS/detailed-balance assumptions, and the dissipative response, diffusion strengths, and memory kernels are not identified beyond what those assumptions justify. For a symmetric Schwarzschild--Schwarzschild thin-shell wormhole, we use the transparent conservative shell to define the reference potential and restoring frequency, and then add damping and stochastic forcing as open-system corrections rather than imposing the conservative first integral on the noisy trajectory. A local constitutive slope $\beta_0^2=(dP/d\sigma)_0$ closes the stable linear response. We show that stability for $2M<a_0<3M$ requires a lower bound on $\beta_0^2$ that diverges as $a_0\to2M^+$, and that the derivative along the family of static junction configurations coincides exactly with the marginal-stability slope; it must therefore not be confused with the constitutive derivative of a stable stochastic shell. The reduced theory yields the stationary distribution, radial and surface-density fluctuations, constitutive pressure and surface-stress fluctuations, quasistatic temperature-scale sensitivities, correlations, Markovian and non-Markovian spectra, and entropy production. In the quasi-horizon regime the junction conditions imply $T_{\rm stress}=T_{\rm acc}+O(\hbar\gamma/a)$, with equality of the leading scales under a precise dominance condition. A peeling temperature joins this relation only when the actual ray-tracing map satisfies an additional generalized Unruh/peeling condition and the usual adiabaticity requirements. The resulting three-temperature correspondence is therefore conditional, while the stochastic coefficients remain phenomenological until derived from a microscopic shell/reservoir model.

\end{abstract}

\maketitle


\section{Introduction}

The discovery that gravitating systems possess genuine thermodynamic properties constitutes one of the deepest developments in modern theoretical physics. The pioneering works of Bekenstein \cite{Bekenstein1972,Bekenstein1973} demonstrated that black holes should carry an entropy proportional to the horizon area, suggesting a profound connection between gravitation, quantum theory and statistical mechanics. Shortly afterwards, Hawking \cite{Hawking1974,Hawking1975} showed that quantum field theory in curved spacetime predicts thermal particle creation by stationary black holes, establishing black holes as genuine thermodynamic systems characterized by a temperature proportional to their surface gravity. Together with the laws of black-hole mechanics formulated by Bardeen, Carter and Hawking \cite{Bardeen1973}, these discoveries established black-hole thermodynamics as a cornerstone of semiclassical gravity.

Numerous independent approaches have since reinforced the intimate relation between gravity and thermodynamics. These include the Euclidean path-integral formulation \cite{Gibbons1977}, the membrane paradigm \cite{Thorne1986}, Jacobson's derivation of Einstein's equations from the Clausius relation \cite{Jacobson1995}, Verlinde's entropic gravity proposal \cite{Verlinde2011}, and Padmanabhan's thermodynamic interpretation of gravitational dynamics \cite{Padmanabhan2010}. These developments strongly suggest that thermodynamics is not merely an analogy but rather an intrinsic aspect of gravitational physics.

The thermodynamic interpretation of gravity naturally raises the question of whether compact objects without event horizons may also possess meaningful thermodynamic descriptions. Traversable wormholes constitute one of the most intriguing solutions of Einstein's equations. Originally introduced by Einstein and Rosen \cite{Einstein1935} as bridges connecting different asymptotically flat regions, their modern interpretation was established by Morris and Thorne \cite{Morris1988,Morris1988b}, who demonstrated that traversable wormholes require exotic matter violating the classical energy conditions. Since then, wormholes have become an important theoretical laboratory for investigating causality, topology change, quantum gravity and modified theories of gravitation \cite{VisserBook,Lobo2017}.

Among the various wormhole constructions, thin-shell wormholes occupy a particularly important position. Instead of specifying a continuous matter distribution throughout the entire spacetime, the exotic matter is confined to an infinitesimally thin hypersurface joining two independent geometries. This construction, based upon the Darmois--Israel junction formalism \cite{Darmois1927,Israel1966}, was systematically developed in the wormhole context by Visser \cite{Visser1989}. Poisson and Visser then established the classic linearized stability analysis for Schwarzschild thin-shell wormholes \cite{PoissonVisser1995}, while the generic dynamical and stability formalism for arbitrary spherically symmetric thin-shell wormholes was developed by Garc\'ia, Lobo, and Visser \cite{Garcia2012}. Thin-shell wormholes provide a remarkably simple framework in which the complete dynamics is encoded in the motion of the throat together with the constitutive properties of the shell.

The dynamics of thin-shell wormholes has been extensively investigated over the last three decades. Linearized stability and generic dynamical thin-shell formalisms were developed in Refs.~\cite{PoissonVisser1995,Garcia2012,Eiroa2008,LoboCrawford2005,Rippentrop2025}, while junction constructions have subsequently been extended to modified-gravity settings, including $F(R)$ gravity \cite{EiroaFigueroa2016}. These works establish the classical mechanical background on which the stochastic construction below is built. The present analysis does not replace the shell equation of state or the junction conservation law by stochastic thermodynamics; rather, it supplements a specified local mechanical closure by phenomenological open-system response and noise.

Thermodynamic descriptions of self-gravitating shells and thin-shell wormholes provide an important complementary line of development. Quasistatic shell thermodynamics was analyzed, for example, in Ref.~\cite{Martinez1996}, while thermodynamic stability has been studied explicitly for Schwarzschild thin-shell wormholes and, more recently, together with dynamical stability for charged thin-shell wormholes \cite{Forghani2019,Eiroa2024}. Recent thin-shell constructions have also combined shell thermodynamics with radial stability in more general environments, including cosmic-void backgrounds \cite{Reboucas2026}. Related notions of entropy, surface gravity, and temperature have been developed for dynamical, trapping, isolated, and apparent horizons \cite{Hayward1998,Hayward1999,Ashtekar2004,Booth2005,Nielsen2009,Faraoni2015}, while self-gravitating shells provide a complementary setting in which no event horizon is required. In addition, Hawking-like particle creation need not rely on a global event horizon: an approximately exponential peeling of null rays can generate an approximately thermal spectrum when the appropriate adiabatic conditions are satisfied \cite{Barcelo2011,Barcelo2011JHEP,Barcelo2006,Visser2003}.

Motivated by these developments, an effective thermodynamic description of dynamical thin-shell wormholes was recently proposed in Ref.~\cite{Lobo:2026vrn}. In that construction, the throat dynamics defines a local acceleration scale and the junction pressure defines a surface-stress scale. Their leading quasi-horizon behavior can coincide, but this geometrical correspondence should not by itself be interpreted as a universal thermodynamic temperature or as evidence for asymptotic particle creation. One objective of the present work is precisely to separate these logically distinct statements.

From a broader perspective, every mesoscopic thermodynamic system is inevitably subject to fluctuations. Ordinary thermodynamics describes only the average macroscopic evolution, whereas microscopic degrees of freedom continuously generate random deviations around the deterministic trajectory. The theoretical framework describing such phenomena is stochastic thermodynamics, whose foundations were established through the pioneering works of Einstein on Brownian motion \cite{Einstein1905}, Langevin's stochastic equation \cite{Langevin1908}, Fokker and Planck's probabilistic formulation \cite{Fokker1914,Planck1917}, Onsager's theory of irreversible processes \cite{Onsager1931a,Onsager1931b}, Kubo's linear response theory \cite{Kubo1966}, and later developments establishing the modern formulation of stochastic thermodynamics \cite{Sekimoto2010,Seifert2005,Seifert2012}.

Stochastic thermodynamics has become a standard theoretical tool for describing mesoscopic systems far from equilibrium, including colloidal particles, nanoscale devices, biological systems and quantum open systems. Central concepts such as Langevin equations, fluctuation--dissipation relations, Fokker--Planck equations, entropy production, fluctuation theorems and nonequilibrium steady states provide a unified description of irreversible processes beyond the traditional equilibrium framework \cite{Risken1989,VanKampen2007,Gardiner2009}.

Stochastic methods have also entered gravitational thermodynamics, although much less systematically than in mesoscopic statistical physics. Langevin and phase-space Fokker--Planck descriptions have been used, for example, to study stochastic black-hole phase-transition kinetics \cite{LiZhangWang2021}, while a recent Fokker--Planck analysis has explicitly tracked probability evolution, Shannon entropy, and entropy production during black-hole thermodynamic transitions \cite{WangEtAl2026}. At the semiclassical level, quantum stress-tensor fluctuations source the Einstein--Langevin equation of stochastic gravity, where noise and dissipative response arise from an open-system treatment of quantum fields \cite{HuVerdaguer2008,CalzettaHu2008,SinhaRavalHu2003}. These developments motivate a corresponding reduced stochastic description of thin-shell degrees of freedom, provided the phenomenological level of that description is made explicit.

Dynamical thin-shell wormholes are particularly suitable for such an investigation. The throat behaves as a finite open system coupled to two asymptotic sectors. In the effective description adopted here these sectors are treated as two gravitational reservoirs. Statistical independence is not assumed as a fundamental property of the global wormhole spacetime: it is an additional approximation, appropriate for factorized or effectively decohered reservoir states. When this approximation is not justified, cross-correlations between the two reservoir sectors must be included explicitly rather than silently discarded. Such a configuration naturally raises questions concerning nonequilibrium transport, stochastic fluctuations and entropy production in gravitational systems.

Building on the effective shell thermodynamics and preliminary stochastic ingredients introduced in Ref.~\cite{Lobo:2026vrn}, the present work develops the open-system structure in a more systematic form. The new emphasis is on the phase-space process $(a,\dot a)$ as the fundamental stochastic description; the explicit overdamped closure required before reducing to a one-variable Fokker--Planck problem; the distinction between the transparent conservative shell and the subsequently added dissipative/noisy response; the separation of constitutive fluctuations from derivatives along a family of static junction solutions; the distinction between probability-current and reservoir entropy production; a dimensionally consistent memory-dependent susceptibility together with colored forcing; and the precise additional assumptions required before a peeling temperature can be compared with the local shell scales.

A further objective is to connect the phenomenological stochastic force with a possible microscopic source in semiclassical gravity. Quantum stress-tensor fluctuations can induce stochastic fluxes on the shell after projection and coarse graining. In stochastic gravity the relevant object is the state-dependent stress-tensor noise kernel \cite{HuVerdaguer2008,PhillipsHu2001}, and a consistent open-system treatment simultaneously generates retarded dissipative response \cite{SinhaRavalHu2003}. Thus the white- and colored-noise models adopted below are effective parametrizations, not microscopic fluctuation--dissipation theorems.

Finally, we specialize the formalism to Schwarzschild--Schwarzschild thin-shell wormholes, first considering asymmetric configurations with independent masses $M_{+}$ and $M_{-}$, and subsequently the symmetric case. Particular attention is devoted to the quasi-horizon regime. We show directly from the junction conditions that the acceleration and surface-stress temperature scales approach one another additively as the quasi-horizon is approached, and that they share the same leading scale whenever the acceleration numerator dominates over the subleading $\gamma^2/a$ contribution. We then identify the additional ray-tracing and adiabaticity conditions under which the semiclassical peeling temperature shares the same asymptotic scale. The resulting three-temperature relation is therefore conditional rather than universal.

The paper is organized as follows. Section~\ref{sec:thermo} combines the two-sided shell thermodynamics with the fundamental phase-space stochastic dynamics. Section~\ref{sec:stochastic} develops the overdamped one-dimensional reduction and then incorporates nonequilibrium two-reservoir transport, non-Markovian memory, spectral response, and conditional noise-induced transitions within a single reduced stochastic framework. Section~\ref{sec:SchSch} applies this construction to Schwarzschild--Schwarzschild thin-shell wormholes and derives the local stability, constitutive fluctuations, quasistatic sensitivities, spectra, and entropy-production sector. Section~\ref{sec:quasi} combines the quasi-horizon expansion with semiclassical particle creation and the conditional unification of the acceleration, surface-stress, and peeling scales. Section~\ref{sec:conclusion} summarizes the results and their domain of validity.

Throughout this work we adopt geometrized units $G=c=k_B=1$, while $\hbar$ is kept explicit whenever quantum effects are considered. Our metric signature is $(-,+,+,+)$.

\section{Effective Shell Thermodynamics and Stochastic Dynamics}
\label{sec:thermo}

In this section we briefly review the effective thermodynamic description of a dynamical thin-shell wormhole connecting two arbitrary static and spherically symmetric universes, establishing the effective quantities that will be required in the stochastic formulation developed in the following sections.

\subsection{Two-sided thin-shell formalism}

We consider two independent four-dimensional spacetimes,
\begin{equation}
ds_{\pm}^{2}
=
-f_{\pm}(r_{\pm})dt_{\pm}^{2}
+
\frac{dr_{\pm}^{2}}
{f_{\pm}(r_{\pm})}
+
r_{\pm}^{2}d\Omega^{2},
\label{metricpm}
\end{equation}
which are joined across a timelike hypersurface $\Sigma: r_{\pm}=a(\tau)$, where $a(\tau)$ denotes the dynamical throat radius and $\tau$ is the proper time measured by an observer comoving with the shell. The induced metric on the throat is $ds_{\Sigma}^{2}=-d\tau^{2}+a^{2}(\tau)d\Omega^{2}$.

Following the standard Darmois--Israel junction formalism \cite{VisserBook,Israel1966,PoissonVisser1995}, the extrinsic curvatures on both sides are $K^{\theta}_{\theta,\pm}=\pm \gamma_{\pm}/a$ and $K^{\tau}_{\tau,\pm}=\pm (\ddot a+\frac12f_{\pm}'(a))/\gamma_{\pm}$, where $\gamma_{\pm}=\sqrt{f_{\pm}(a)+\dot a^{2}}$. The Israel equations $S^{i}_{\ j}=-(1/8\pi)([K^{i}_{\ j}]-\delta^{i}_{\ j}[K])$ yield the surface energy density
\begin{equation}
\sigma = -\frac1{4\pi a}(\gamma_{+}+\gamma_{-}),
\label{sigma}
\end{equation}
and the surface pressure
\begin{equation}
P = \frac1{8\pi}\left[
\frac{\ddot a+\frac12f_{+}'(a)}{\gamma_{+}}
+
\frac{\ddot a+\frac12f_{-}'(a)}{\gamma_{-}}
+
\frac{\gamma_{+}+\gamma_{-}}{a}
\right].
\label{pressure}
\end{equation}

These expressions remain valid for completely asymmetric configurations, $f_{+}\neq f_{-}$, and naturally reduce to the usual symmetric thin-shell wormhole when both geometries coincide.

Assuming that no external matter crosses the shell, the intrinsic conservation law $S^{i}_{\ j|i}=0$ becomes the first-law form
\begin{equation}
dU+PdA=0,
\end{equation}
where $A=4\pi a^{2}$ is the shell area and $U=\sigma A$ is the internal energy. Therefore, in the absence of external energy fluxes, the throat evolves adiabatically.

The equation above is the \emph{transparent conservative limit}. Once mean exchange with the two environments is introduced, the shell balance instead takes the oriented open-system form $dU/d\tau+P\,dA/d\tau=J_+-J_-$ used in Sec.~\ref{sec:nonEq}, with the general thin-shell flux structure discussed in Ref.~\cite{LoboCrawford2005}, and with stochastic fluctuations superposed on the mean currents. Accordingly, the conservative junction potential introduced below is used as a reference mechanical potential that fixes the local restoring stiffness. The dissipative and stochastic forces are additional open-system terms; after they are switched on, the conservative first integral is not imposed as an exact constraint on every noisy trajectory.

\subsection{Effective temperatures}

The local dynamics of the throat naturally defines an effective acceleration measured independently from each bulk geometry \cite{Lobo:2026vrn}:
\begin{equation}
\kappa_{\pm}
=
\frac{
\left|
\ddot a+\frac12f_{\pm}'(a)
\right|
}
{\sqrt{f_{\pm}(a)+\dot a^{2}}}.
\label{kappaside}
\end{equation}

Since the shell connects two independent universes, neither side is privileged. A natural effective quantity is therefore obtained through the arithmetic average
\begin{equation}
\kappa_{\rm eff}
=
\frac12
\left(
\kappa_{+}
+
\kappa_{-}
\right).
\label{kappaeff}
\end{equation}

The corresponding effective Unruh-like acceleration temperature is motivated by the standard acceleration-temperature relation \cite{Unruh1976}:
\begin{equation}
T_{\rm acc}
=
\frac{\hbar}{2\pi}
\kappa_{\rm eff}.
\label{Tacc}
\end{equation}

Unlike the Hawking temperature of a stationary event horizon, this quantity is a local Unruh-like acceleration scale associated with the proper acceleration required to maintain the shell trajectory. For a genuinely time-dependent trajectory in curved spacetime, a particle detector need not exhibit an exactly Planckian response at this instantaneous value; such a thermal interpretation additionally requires an appropriate adiabatic/local-stationarity regime. The acceleration scale itself remains well defined even in the absence of horizons.

An alternative characterization follows from the isotropic tangential surface stress of the shell. With the convention $S^{i}{}_{j}={\rm diag}(-\sigma,P,P)$, $P>0$ denotes tangential pressure while $P<0$ corresponds to mechanical tension $-P$. Since the quantity entering the effective scale is the magnitude $|P|$, it is more precise to refer to it as a surface-stress scale rather than as a mechanical tension in all regimes. We therefore define phenomenologically
\begin{equation}
\kappa_{\rm stress}=4\pi |P|,
\label{kappastress}
\end{equation}
with corresponding temperature
\begin{equation}
T_{\rm stress}
=
\frac{\hbar}{2\pi}
\kappa_{\rm stress}.
\label{Tstress}
\end{equation}

Unlike the acceleration temperature, which depends explicitly on the shell trajectory, the surface-stress temperature scale is directly associated with the local surface stresses determined by the junction conditions. In general, both temperatures are independent quantities. As shown in Sec.~\ref{sec:unification}, their difference is bounded by a term of order $\gamma$ in the symmetric quasi-horizon regime, and they share the same leading scale under an additional dominance condition. Thus the geometrical correspondence is asymptotic and conditional rather than an exact identity at finite radius.

The quantities introduced above define the deterministic thermodynamic sector of the throat. The shell dynamics determines the effective acceleration temperature through Eq.~(\ref{Tacc}), while the local stresses determine the effective surface-stress temperature scale through Eq.~(\ref{Tstress}). These quantities are macroscopic effective scales associated with the throat. Their geometrical definitions do not require a microscopic statistical model, but a literal thermodynamic-temperature interpretation requires additional assumptions specified below.

We now promote this deterministic shell description to an effective stochastic one by allowing the radial trajectory to fluctuate about its mean evolution. The phase-space process introduced next is the fundamental stochastic level of the construction; the effective temperature scales defined above enter only as phenomenological inputs to noise and transport coefficients when the additional local-equilibrium assumptions stated below are imposed.


\subsection{Phase-Space Stochastic Dynamics of the Throat}
\label{sec:phaseSpaceStochastic}

The Israel junction conditions show that the surface variables are not independent
mechanical coordinates.  In the general two-sided geometry one has
$\sigma=\sigma(a,\dot a)$, whereas $P=P(a,\dot a,\ddot a)$.  It is therefore
natural to formulate the stochastic dynamics first in terms of the geometrical
degree of freedom of the shell, namely its radius and radial velocity, and only
afterwards regard $\sigma$, $P$, $T_{\rm acc}$ and $T_{\rm stress}$ as stochastic
observables induced by the random trajectory.  This phase-space formulation is
the fundamental stochastic description adopted in this section.  The
one-dimensional description introduced in the next section will be obtained only
after an overdamped/adiabatic reduction.

\subsubsection{Radial phase space and deterministic drift}

We introduce
\begin{equation}
v(\tau)\equiv\dot a(\tau),
\end{equation}
and write the deterministic radial evolution in the general first-order form
\begin{equation}
\dot a=v,\qquad
\dot v=F(a,v).
\label{DeterministicPhaseSpace}
\end{equation}
The function $F(a,v)$ summarizes the deterministic shell dynamics after an
equation of state, constitutive law and mean exchange with the exterior sectors
have been specified.  For a conservative thin shell whose radial dynamics may be
written as
\begin{equation}
\dot a^2+V(a)=0,
\label{RadialPotential}
\end{equation}
differentiation along a nonturning trajectory gives
\begin{equation}
\ddot a=-\frac12 V'(a),
\end{equation}
so that $F(a,v)=-V'(a)/2$.  More generally, an open shell may contain dissipative
terms and one may write, locally,
\begin{equation}
F(a,v)=-\frac12V'(a)-\Upsilon(a,v)v+F_{\rm ext}(a,v),
\label{GeneralRadialDrift}
\end{equation}
where $\Upsilon$ is an effective radial damping coefficient and $F_{\rm ext}$
contains the mean force induced by the environments.

In Eq.~(\ref{GeneralRadialDrift}), $V(a)$ should therefore be read as the conservative \emph{reference} potential. If $\Upsilon$, $F_{\rm ext}$, or stochastic forcing are nonzero, Eq.~(\ref{RadialPotential}) is not simultaneously an exact first integral of the open dynamics. This separation is used explicitly in the Schwarzschild worked example.

The junction variables are then observables on this stochastic phase space:
\begin{equation}
\sigma(a,v)
=
-\frac{1}{4\pi a}
\left[
\sqrt{f_+(a)+v^2}+\sqrt{f_-(a)+v^2}
\right],
\label{SigmaPhaseSpace}
\end{equation}
and
\begin{eqnarray}
P(a,v,\dot v)
&=&
\frac{1}{8\pi}
\Bigg[
\frac{\dot v+\frac12f_+'(a)}{\sqrt{f_+(a)+v^2}}
+
\frac{\dot v+\frac12f_-'(a)}{\sqrt{f_-(a)+v^2}}
\nonumber\\
&&+
\frac{\sqrt{f_+(a)+v^2}+\sqrt{f_-(a)+v^2}}{a}
\Bigg].
\label{PressurePhaseSpace}
\end{eqnarray}
Thus a stochastic trajectory $(a(\tau),v(\tau))$ automatically generates
stochastic trajectories for the surface energy density and surface pressure.

\subsubsection{Two-reservoir stochastic forcing}

The two asymptotic sectors act on the shell through fluctuating stress-energy
fluxes.  At the microscopic level the random input may be represented by
$\delta\Phi_\pm$.  A radial Langevin equation requires the projection of these
flux fluctuations onto the radial mechanical degree of freedom.  We therefore
introduce response/projection coefficients ${\cal C}_\pm(a,v)$ and define the
effective radial random acceleration
\begin{equation}
\eta(\tau)
=
{\cal C}_+(a,v)\,\delta\Phi_+(\tau)
+
{\cal C}_-(a,v)\,\delta\Phi_-(\tau).
\label{RadialNoiseProjection}
\end{equation}
The coefficients ${\cal C}_\pm$ are not fixed by the conservation equation
alone; microscopically they would follow from the retarded shell response to the
projected stress-tensor fluctuations.  This distinction prevents the energy-flux
noise from being identified, without justification, with a radial force.

The stochastic phase-space equations are
\begin{equation}
da=v\,d\tau,
\label{SDEa}
\end{equation}
\begin{equation}
dv=F(a,v)\,d\tau+\sqrt{2D_v(a,v)}\,dW_\tau,
\label{SDEv}
\end{equation}
where $W_\tau$ is a Wiener process.  Equivalently,
$\eta(\tau)=\sqrt{2D_v}\,\dot W_\tau$ and
\begin{equation}
\langle\eta(\tau)\eta(\tau')\rangle
=
2D_v(a,v)\delta(\tau-\tau').
\label{RadialWhiteNoise}
\end{equation}
For statistically independent reservoirs the local diffusion strength is the
sum of the two projected contributions,
\begin{equation}
D_v=D_{v,+}+D_{v,-}.
\label{DvSum}
\end{equation}
If the reservoir states are correlated, a cross-diffusion contribution must be
added.

Under the same local-thermal/KMS and local-detailed-balance assumptions used
below, a phenomenological two-reservoir Einstein parametrization may be written
as
\begin{equation}
D_v=\mu_+T_+ +\mu_-T_-,
\label{RadialEinstein}
\end{equation}
where $\mu_\pm\ge0$ are radial mobility-response coefficients with the dimensions
required by Eq.~(\ref{SDEv}); their non-negativity makes the positivity of the
corresponding diffusion contributions explicit when $T_\pm>0$.  Equation~(\ref{RadialEinstein}) is not a
microscopic fluctuation--dissipation theorem unless $\mu_\pm$ and the
dissipative part of $F$ are derived from the same reservoir couplings.

\subsubsection{Kramers equation}

Let $\rho(a,v,\tau)$ denote the probability density in radial phase space.  In
the It\^o convention, Eqs.~(\ref{SDEa})--(\ref{SDEv}) imply
\begin{equation}
\frac{\partial\rho}{\partial\tau}
=
-\frac{\partial}{\partial a}(v\rho)
-\frac{\partial}{\partial v}\!\left[F(a,v)\rho\right]
+
\frac{\partial^2}{\partial v^2}\!\left[D_v(a,v)\rho\right].
\label{KramersPhaseSpace}
\end{equation}
For constant $D_v$ the last term reduces to
$D_v\partial_v^2\rho$.  Equation~(\ref{KramersPhaseSpace}) is the natural
Markovian stochastic equation for a genuinely dynamical thin shell.  It retains
the distinction between radial position and velocity and does not require
$\sigma$ to be an independent stochastic coordinate.

\subsubsection{Linear fluctuations about a stationary throat}

Consider a stationary radius $a_0$ with $v_0=0$ and
$F(a_0,0)=0$.  Define
\begin{equation}
x=\delta a=a-a_0,\qquad u=\delta v=v.
\end{equation}
To linear order,
\begin{equation}
\dot x=u,\qquad
\dot u=-\omega_0^2x-\gamma_0u+\eta(\tau),
\label{LinearRadialLangevin}
\end{equation}
where
\begin{equation}
\omega_0^2
=
-\left(\frac{\partial F}{\partial a}\right)_0,
\qquad
\gamma_0
=
-\left(\frac{\partial F}{\partial v}\right)_0.
\label{omegaGamma}
\end{equation}
For the conservative equation $\dot a^2+V(a)=0$,
\begin{equation}
\omega_0^2=\frac12V''(a_0).
\end{equation}
Thus the usual thin-shell stability requirement $V''(a_0)>0$ becomes
$\omega_0^2>0$.  A stationary stochastic state additionally requires
$\gamma_0>0$.

For additive white noise
$\langle\eta(\tau)\eta(\tau')\rangle=2D_{v0}\delta(\tau-\tau')$, the stationary
covariances of the damped linear system are
\begin{equation}
\langle(\delta v)^2\rangle
=
\frac{D_{v0}}{\gamma_0},
\qquad
\langle\delta a\,\delta v\rangle=0,
\label{VelocityVariance}
\end{equation}
and
\begin{equation}
\langle(\delta a)^2\rangle
=
\frac{D_{v0}}{\gamma_0\omega_0^2}.
\label{RadialVariance}
\end{equation}
These relations provide the direct bridge between mechanical stability and
stochastic radial fluctuations.

\subsubsection{Induced fluctuations of the surface observables}

For any smooth shell observable ${\cal O}(a,v)$, its linear fluctuation is
\begin{equation}
\delta{\cal O}
=
{\cal O}_{,a}\,\delta a
+
{\cal O}_{,v}\,\delta v.
\end{equation}
Hence
\begin{eqnarray}
\langle(\delta{\cal O})^2\rangle
&=&
{\cal O}_{,a}^2\langle(\delta a)^2\rangle
+
{\cal O}_{,v}^2\langle(\delta v)^2\rangle
\nonumber\\
&&+
2{\cal O}_{,a}{\cal O}_{,v}
\langle\delta a\,\delta v\rangle.
\label{ObservableVariance}
\end{eqnarray}
For $\sigma(a,v)$, Eq.~(\ref{SigmaPhaseSpace}) is even in $v$, so that
$\sigma_{,v}|_{v=0}=0$.  Therefore, about a static throat,
\begin{equation}
\langle(\delta\sigma)^2\rangle
=
\left(\frac{\partial\sigma}{\partial a}\right)_0^2
\frac{D_{v0}}{\gamma_0\omega_0^2}.
\label{SigmaVarianceFromRadial}
\end{equation}
The pressure contains the acceleration and consequently its instantaneous
white-noise fluctuation is distribution-valued if Eq.~(\ref{SDEv}) is used
literally.  A finite pressure variance therefore requires either a finite
coarse-graining time, colored noise, or the reduced overdamped description
introduced below.  This is a physical ultraviolet limitation of the white-noise
idealization rather than a peculiarity of the junction formalism.

\section{Reduced Stochastic Thermodynamics, Nonequilibrium Transport, and Memory}
\label{sec:stochastic}

The phase-space description above is the appropriate starting point for the
complete radial dynamics.  A one-dimensional stochastic theory is nevertheless
useful when the velocity relaxes much faster than the radius, or when the
dynamics is restricted to a one-parameter quasistatic branch.  We now impose
precisely this reduction and require that the stochastic state be described by
the throat radius alone.

\subsection{Overdamped/adiabatic closure}

Assume that the fast variables can be adiabatically eliminated so that, on the
coarse-graining time scale,
\begin{equation}
\dot a={\cal A}(a)+\zeta(\tau),
\label{ReducedLangevinA}
\end{equation}
with
\begin{equation}
\langle\zeta(\tau)\rangle=0,\qquad
\langle\zeta(\tau)\zeta(\tau')\rangle
=
2D_a\,\delta(\tau-\tau').
\label{ReducedNoiseA}
\end{equation}
The closure assumption is therefore
\begin{equation}
{\cal A}={\cal A}(a),\qquad D_a={\rm const.}
\label{closure}
\end{equation}
locally in the relevant radial interval.  The coefficient $D_a$ is the
diffusion coefficient of the \emph{reduced radial process}; it should not be
identified with the energy-flux diffusion coefficient or with $D_v$ without
performing the appropriate response projection and adiabatic elimination.

If the eliminated velocity obeys locally
$\dot v\simeq-\gamma(a)v+F_0(a)+\eta$, with $\gamma$ large compared with the
radial evolution rate, then
\begin{equation}
{\cal A}(a)\simeq\frac{F_0(a)}{\gamma(a)},
\qquad
D_a(a)\simeq\frac{D_v(a)}{\gamma^2(a)}.
\label{AdiabaticReduction}
\end{equation}
Equation~(\ref{closure}) corresponds to taking these coefficients approximately
constant or radius-dependent only over the local domain of interest. For genuinely state-dependent damping or diffusion, a systematic
adiabatic elimination can generate multiplicative-noise and noise-induced-drift terms whose precise form depends on the stochastic
convention and microscopic reduction. These terms are outside the local constant-coefficient closure adopted here and should not be
silently absorbed into Eq.~(\ref{AdiabaticReduction}).

\subsection{Two-reservoir noise scale}

Under local thermal/KMS conditions, the reduced diffusion strength may be
parametrized phenomenologically as
\begin{equation}
D_a=\Gamma_{a,+}T_+ +\Gamma_{a,-}T_-,
\label{FDTcorrected}
\end{equation}
where $\Gamma_{a,\pm}\ge0$ are reduced radial mobilities, so that the
phenomenological diffusion strength is non-negative for positive bath
temperatures.  For equal couplings,
$\Gamma_{a,+}=\Gamma_{a,-}=\Gamma_{a,\rm eff}$,
\begin{equation}
D_a=2\Gamma_{a,\rm eff}T_{\rm acc}.
\label{FDTsymmetric}
\end{equation}
As before, this becomes a genuine fluctuation--dissipation relation only if the
same microscopic couplings generate both the reduced dissipative drift and the
noise.

\subsection{Linear radial fluctuations}

Let $a_0$ be a stable fixed point of the reduced dynamics,
${\cal A}(a_0)=0$, and write $a=a_0+\delta a$.  To first order,
\begin{equation}
\frac{d(\delta a)}{d\tau}
=
-\lambda_a\delta a+\zeta(\tau),
\label{OU}
\end{equation}
where
\begin{equation}
\lambda_a
=
-\left(\frac{d{\cal A}}{da}\right)_0>0.
\label{lambdadef}
\end{equation}
The stationary correlation and variance are
\begin{equation}
\langle\delta a(\tau)\delta a(0)\rangle
=
\frac{D_a}{\lambda_a}e^{-\lambda_a|\tau|},
\label{correlation}
\end{equation}
\begin{equation}
\langle(\delta a)^2\rangle=\frac{D_a}{\lambda_a}.
\label{variance}
\end{equation}
For a one-parameter branch $\sigma=\sigma(a)$ and $P=P(a)$,
\begin{equation}
\langle(\delta\sigma)^2\rangle
=
[\sigma'(a_0)]^2\frac{D_a}{\lambda_a},\quad
\langle(\delta P)^2\rangle
=
[P'(a_0)]^2\frac{D_a}{\lambda_a}.
\label{ReducedSurfaceVariances}
\end{equation}

Here $P'(a_0)$ denotes the derivative prescribed by the chosen reduced constitutive branch. It must not be replaced automatically by the derivative obtained by moving through a family of distinct static junction configurations. The latter is a geometrical static-locus derivative and, as shown explicitly in Sec.~\ref{sec:SchClosure}, it coincides with the marginal-stability constitutive slope in the symmetric Schwarzschild example.

\subsection{Fokker--Planck equation}

Let ${\cal P}(a,\tau)$ denote the probability density for the throat radius.
Equations~(\ref{ReducedLangevinA})--(\ref{ReducedNoiseA}) give
\begin{equation}
\frac{\partial{\cal P}}{\partial\tau}
=
-\frac{\partial}{\partial a}
\left[{\cal A}(a){\cal P}\right]
+
D_a\frac{\partial^2{\cal P}}{\partial a^2}.
\label{FP}
\end{equation}
The probability current is
\begin{equation}
J_a={\cal A}(a){\cal P}-D_a\frac{\partial{\cal P}}{\partial a}.
\label{probcurrent}
\end{equation}
For a stationary zero-current state, provided the normalization integral exists and the chosen boundary conditions are compatible with zero probability current,
\begin{equation}
{\cal P}_{\rm st}(a)
=
{\cal N}
\exp\left[
\int^a\frac{{\cal A}(x)}{D_a}\,dx
\right].
\label{Peq}
\end{equation}
Defining the radial stochastic quasipotential
\begin{equation}
{\cal V}_{a}(a)=-\int^a{\cal A}(x)\,dx,
\label{potential}
\end{equation}
one obtains
\begin{equation}
{\cal P}_{\rm st}(a)
=
{\cal N}\exp\left[-\frac{{\cal V}_a(a)}{D_a}\right].
\label{Boltzmann}
\end{equation}
Near a stable fixed point,
${\cal V}_a\simeq{\cal V}_0+\lambda_a(a-a_0)^2/2$, and the stationary density is
Gaussian with the variance in Eq.~(\ref{variance}).

\subsection{Coarse-grained stochastic entropy}

Using a constant reference density ${\cal P}_*$, define
\begin{equation}
S_{\rm cg}
=
-\int {\cal P}(a,\tau)
\ln\!\left[\frac{{\cal P}(a,\tau)}{{\cal P}_*}\right]da.
\end{equation}
For vanishing boundary terms,
\begin{equation}
\frac{dS_{\rm cg}}{d\tau}
=
\Pi_{\rm cg}-\Phi_{S,\rm cg},
\end{equation}
where
\begin{equation}
\Pi_{\rm cg}
=
\int\frac{J_a^2}{D_a{\cal P}}\,da\ge0,
\label{Pi}
\end{equation}
and
\begin{equation}
\Phi_{S,\rm cg}
=
\int\frac{{\cal A}J_a}{D_a}\,da.
\end{equation}
The vanishing of $J_a$ establishes detailed balance only for the reduced radial
coordinate; hidden phase-space or reservoir currents may remain nonzero.

\subsection{Domain of validity}

The reduction to Eq.~(\ref{ReducedLangevinA}) requires a separation of time
scales, a one-parameter quasistatic branch, or an explicitly overdamped radial
response.  Outside this regime the Kramers equation
(\ref{KramersPhaseSpace}), rather than the one-dimensional Fokker--Planck
equation (\ref{FP}), is the appropriate stochastic description.  In
particular, close to a quasi-horizon the effective temperatures and response
coefficients may vary rapidly with $a$; the constant-$D_a$ approximation must
then be understood locally and may eventually fail.

\subsection{Nonequilibrium thermodynamics of two gravitational reservoirs}
\label{sec:nonEq}

The throat may exchange energy with two asymptotic environments. Each side also defines the local acceleration scale
\begin{equation}
T_{\pm}=\frac{\hbar\kappa_{\pm}}{2\pi}.
\end{equation}
It is essential, however, to distinguish this observer-dependent Unruh-like quantity from the temperature of a genuine thermal reservoir. In the discussion below we adopt a local-equilibrium approximation: the quantum or microscopic state coupled to the shell is assumed to be approximately KMS, or at least to satisfy local detailed balance, at the corresponding effective scale. Only under this additional assumption may $T_\pm$ be used as bath temperatures in an irreversible-thermodynamic description.

When the two effective reservoir temperatures differ, define $\Delta T=T_+-T_-$. Since
$T_{\rm acc}=(T_++T_-)/2$ by Eqs.~(\ref{kappaeff})--(\ref{Tacc}), the natural reference temperature close to equilibrium is already the acceleration scale introduced above. For $|\Delta T|\ll T_{\rm acc}$, define the thermodynamic affinity
\begin{equation}
{\cal X}_T
=
\frac{1}{T_-}-\frac{1}{T_+}
\simeq
\frac{\Delta T}{T_{\rm acc}^2}.
\label{Affinity}
\end{equation}

To avoid sign ambiguities, we use oriented heat currents. Let $J_+$ denote positive energy flow from the $+$ environment into the shell and let $J_-$ denote positive energy flow from the shell into the $-$ environment. The shell energy balance is then
\begin{equation}
\frac{dU}{d\tau}
=
J_+-J_- -P\frac{dA}{d\tau}.
\label{orientedbalance}
\end{equation}
In a steady state with negligible energy storage and mechanical work, $J_+=J_-\equiv J_Q$, and the same $J_Q$ is the heat current transmitted through the throat. This oriented definition avoids the factor-of-two ambiguity that arises if both side fluxes are defined as positive when entering the shell.

The reservoir entropy production associated with this steady heat transfer is
\begin{equation}
\Pi_Q
=
J_Q
\left(
\frac1{T_-}-\frac1{T_+}
\right).
\label{EntropyHeat}
\end{equation}
Close to equilibrium, Onsager theory gives
\begin{equation}
J_Q=L_Q{\cal X}_T,
\label{Onsager}
\end{equation}
with $L_Q\ge0$, so that
\begin{equation}
\Pi_Q=L_Q{\cal X}_T^2\ge0.
\label{PositiveEntropy}
\end{equation}

These relations are valid within the local-thermal/detailed-balance approximation stated above; the mere inequality $\kappa_+\neq\kappa_-$ is not by itself sufficient to prove the existence of a heat current.

The stochastic fluxes fluctuate around the mean oriented currents. The same microscopic couplings that generate dissipation may then generate fluctuations, and the phenomenological relation (\ref{FDTcorrected}) supplies the reduced noise strength whenever the local-equilibrium approximation is valid. If the asymptotic quantum state is nonthermal, strongly entangled between the two sides, or far from local detailed balance, the transport coefficients and the noise kernel must instead be derived from the corresponding nonequilibrium correlation functions.

\subsection{Non-Markovian stochastic dynamics}
\label{sec:nonmarkov}

The white-noise approximation is appropriate only when the microscopic correlation time is much shorter than the dynamical time of the reduced radial variable. Generalized Langevin descriptions with memory are a standard consequence of eliminating unresolved degrees of freedom \cite{Zwanzig1961}. Curved-spacetime quantum fields, vacuum polarization, and backreaction need not be delta-correlated and can generate temporally nonlocal response. After coarse graining, this naturally motivates an effective non-Markovian description with a finite memory scale.

We introduce the normalized even correlation profile
\begin{equation}
k_c(\Delta\tau)
=
\frac{1}{\tau_c}\exp\left(-\frac{|\Delta\tau|}{\tau_c}\right),
\label{normalizedkernel}
\end{equation}
which satisfies $k_c(\Delta\tau)\to2\delta(\Delta\tau)$ as $\tau_c\to0$ on the full time axis. Here $\tau_c>0$ denotes the phenomenological correlation time. The colored-noise correlation is written
\begin{equation}
\langle\zeta(\tau)\zeta(\tau')\rangle
=D_a\,k_c(\tau-\tau'),
\label{OUColored}
\end{equation}
so that the white-noise limit reproduces Eq.~(\ref{ReducedNoiseA}).

For the linearized reduced dynamics, memory is introduced through
\begin{equation}
\dot{\delta a}(\tau)
=
-\lambda_{a0}\delta a(\tau)
-
\int_0^\tau {\cal M}(\tau-s)\delta a(s)\,ds
+\zeta(\tau),
\label{GLE}
\end{equation}
where $\lambda_{a0}$ is the instantaneous relaxation rate and ${\cal M}$ is a causal memory kernel.

Equation~(\ref{GLE}) is written in initial-value form for a process prepared at $\tau=0$. The spectral analysis in Sec.~\ref{sec:spectrum}, however, refers to the stationary long-time regime, after all preparation-dependent and dynamical relaxation transients have decayed. Equivalently, for the stationary process the causal memory term may be written as
\[
\int_{-\infty}^{\tau}{\cal M}(\tau-s)\,\delta a(s)\,ds.
\]
For stable relaxation and a decaying memory kernel this is the long-time stationary form of Eq.~(\ref{GLE}). Thus $\tau$ must be large compared not only with the memory time $\tau_c$ but also with the slowest relaxation time associated with the poles of the response function.

To keep the dissipative memory sector distinct from the mobility coefficients $\Gamma_{a,\pm}$ entering the Markovian Einstein ansatz, we introduce independent memory amplitudes $\Lambda_\pm$. For exponential kernels,
\begin{equation}
{\cal M}(t)={\cal M}_+(t)+{\cal M}_-(t)\;,\;
{\cal M}_\pm(t)=\Lambda_\pm k_+(t),
\label{KernelSides}
\end{equation}
with
\begin{equation}
k_+(t)=\frac{1}{\tau_c}e^{-t/\tau_c}\Theta(t)\,,\qquad
\int_0^\infty k_+(t)dt=1.
\label{Kernel}
\end{equation}
Hence
\begin{equation}
{\cal M}(t)=\Lambda_{\rm tot}k_+(t)\,,\qquad
\Lambda_{\rm tot}\equiv\Lambda_++\Lambda_-.
\label{KernelTotal}
\end{equation}
For notational economy we use the same correlation time $\tau_c$ in the colored-noise and exponential-memory ansatz. This is a phenomenological simplification, not a microscopic identity; a derived open-system model may contain distinct noise and response correlation times.
The generalized Langevin equation requires $[{\cal M}]={\rm time}^{-2}$; since $[k_+]={\rm time}^{-1}$, the amplitudes $\Lambda_\pm$ have dimensions of an inverse time. They are therefore not dimensionally interchangeable with the mobility-like coefficients $\Gamma_{a,\pm}$ in Eq.~(\ref{FDTcorrected}). In the short-memory limit ${\cal M}(t)\to\Lambda_{\rm tot}\delta_+(t)$, where $\delta_+$ is the one-sided delta appropriate to the causal convolution.

With the normalization in Eq.~(\ref{Kernel}), this limit reduces Eq.~(\ref{GLE}) to
\begin{equation}
\dot{\delta a}=-(\lambda_{a0}+\Lambda_{\rm tot})\delta a+\zeta,
\label{ShortMemoryGLE}
\end{equation}
so continuous recovery of the Markovian Ornstein--Uhlenbeck equation (\ref{OU}) requires
\begin{equation}
\lambda_a=\lambda_{a0}+\Lambda_{\rm tot}.
\label{MarkovianMatching}
\end{equation}
For the exponential kernel, introducing the auxiliary memory variable
$y(\tau)=\int_{-\infty}^{\tau}k_+(\tau-s)\delta a(s)\,ds$ gives the deterministic characteristic equation
\begin{equation}
\tau_c r^2+(1+\lambda_{a0}\tau_c)r+(\lambda_{a0}+\Lambda_{\rm tot})=0.
\label{MemoryCharacteristic}
\end{equation}
A stationary spectrum therefore presupposes that both roots have negative real parts; for $\tau_c>0$ this is equivalent to
$1+\lambda_{a0}\tau_c>0$ and $\lambda_{a0}+\Lambda_{\rm tot}>0$, and is automatically satisfied in the common dissipative case $\lambda_{a0}>0$ and $\Lambda_{\rm tot}\ge0$.

Under the same local-equilibrium assumptions used in the Markovian sector, we may retain the phenomenological colored-noise parametrization
\begin{equation}
\langle\zeta(\tau)\zeta(\tau')\rangle
=
\left(\Gamma_{a,+}T_+ + \Gamma_{a,-}T_-\right)k_c(\tau-\tau').
\label{GeneralFDT}
\end{equation}
Its white-noise limit is
\begin{equation}
\langle\zeta(\tau)\zeta(\tau')\rangle
\rightarrow
2\left(\Gamma_{a,+}T_+ + \Gamma_{a,-}T_-\right)\delta(\tau-\tau'),
\end{equation}
which reproduces $D_a=\Gamma_{a,+}T_++\Gamma_{a,-}T_-$. No identification between $\Lambda_\pm$ and $\Gamma_{a,\pm}$ is made at the phenomenological level. A genuine generalized fluctuation--dissipation theorem would instead derive both the retarded memory kernel and the symmetrized noise kernel from the same microscopic reservoir correlation functions, thereby fixing the relation between their amplitudes and frequency dependence.

Equation~(\ref{GLE}) is thus an effective generalized Langevin equation with independently parametrized memory and noise sectors. A microscopic stochastic-gravity or open-quantum-system derivation would replace these kernels by the appropriate retarded response functions and symmetrized stress-tensor correlators in the specified quantum state.

\subsection{Spectral properties of the stochastic fluctuations}
\label{sec:spectrum}

For the Markovian Ornstein--Uhlenbeck equation (\ref{OU}) \cite{UhlenbeckOrnstein1930},
\begin{equation}
(-i\omega+\lambda_a)\widetilde{\delta a}(\omega)
=
\tilde\zeta(\omega),
\end{equation}
and the susceptibility is
\begin{equation}
\chi_{\rm M}(\omega)
=
\frac1{\lambda_a-i\omega}.
\label{susceptibility}
\end{equation}
Hence
\begin{equation}
S_a(\omega)
=
|\chi_{\rm M}(\omega)|^2S_\zeta(\omega).
\label{SpectralDensity}
\end{equation}
For white noise, $S_\zeta=2D_a$ and
\begin{equation}
S_a^{\rm M}(\omega)
=
\frac{2D_a}{\lambda_a^2+\omega^2},
\label{Lorentzian}
\end{equation}
the usual Lorentzian spectrum.

Accordingly, the non-Markovian Fourier analysis below is performed in the stationary long-time regime, using the time-translation-invariant causal convolution extending from $-\infty$ to $\tau$. The resulting expressions describe stationary spectra, not the finite-time transient response immediately after $\tau=0$.

The generalized Langevin equation has a different response function. Fourier transforming the stationary long-time form described above gives
\begin{equation}
\widetilde{\delta a}(\omega)
=
\chi_{\rm NM}(\omega)\tilde\zeta(\omega),
\end{equation}
with
\begin{equation}
\chi_{\rm NM}(\omega)
=
\frac1{
\lambda_{a0}-i\omega+\widetilde{\cal M}(\omega)
}.
\label{susceptibilityNM}
\end{equation}
Here $\widetilde{\cal M}(\omega)=\int_0^\infty {\cal M}(t)e^{i\omega t}\,dt$ is the one-sided Fourier transform of the causal memory kernel, consistent with the convention for which $d/d\tau\mapsto-i\omega$. For the exponential total causal kernel (\ref{KernelTotal}),
\begin{equation}
\widetilde{\cal M}(\omega)
=
\frac{\Lambda_{\rm tot}}{1-i\omega\tau_c},
\end{equation}
while the colored-noise correlation (\ref{OUColored}) gives
\begin{equation}
S_\zeta(\omega)
=
\frac{2D_a}{1+\omega^2\tau_c^2}.
\end{equation}
Therefore the consistent non-Markovian spectrum is
\begin{equation}
S_a^{\rm NM}(\omega)
=
\frac{2D_a}
{1+\omega^2\tau_c^2}
\left|
\lambda_{a0}-i\omega+
\frac{\Lambda_{\rm tot}}{1-i\omega\tau_c}
\right|^{-2}.
\label{SpectrumNM}
\end{equation}

Thus colored noise and memory modify different pieces of the spectrum: the former changes $S_\zeta$, while the latter changes the susceptibility itself. Keeping the Markovian susceptibility while modifying only $S_\zeta$ describes colored forcing of a Markovian system, not a complete generalized Langevin dynamics.

As a consistency check, when $\tau_c\to0$ one has $S_\zeta\to2D_a$ and $\widetilde{\cal M}\to\Lambda_{\rm tot}$. Using the matching relation (\ref{MarkovianMatching}), Eq.~(\ref{SpectrumNM}) then reduces exactly to the Lorentzian spectrum (\ref{Lorentzian}).

\subsection{Noise-induced transitions in the reduced throat dynamics}
\label{sec:phase}

Within the one-dimensional radial closure (\ref{closure}), a zero-current stationary state defines the stochastic quasipotential
\begin{equation}
{\cal V}_{a}(a)
=
-\int^a {\cal A}(x)\,dx.
\end{equation}
The formalism therefore permits stochastic transitions whenever the reduced drift develops multiple stable fixed points. This statement is conditional: the existence of such minima must be demonstrated for a specified shell equation of state, flux model and closure, and does not follow from the junction conditions alone.

Let $a_c$ be a stationary radius and define $\psi=a-a_c$. Locally,
\begin{equation}
{\cal V}_{a}
=
{\cal V}_c+\frac{a_2}{2}\psi^2
+\frac{a_3}{3!}\psi^3
+\frac{a_4}{4!}\psi^4+\cdots.
\label{Expansion}
\end{equation}
When a $\psi\to-\psi$ symmetry is present, the odd coefficients vanish. A change in sign of $a_2$ with $a_4>0$ then has the familiar Landau form of a continuous bifurcation. When the symmetry is absent, multiple local minima may coexist and noise can induce transitions between them.

In the weak-noise, overdamped, Markovian regime, with a quasipotential barrier $\Delta {\cal V}_{a}$ large compared with the effective noise scale, the escape rate has the Kramers form \cite{Kramers1940}
\begin{equation}
\begin{aligned}
\Gamma_{\rm esc}
&\simeq
\Gamma_0\exp\left(-\frac{\Delta {\cal V}_{a}}{D_a}\right)\\
&=
\Gamma_0\exp\left[
-\frac{\Delta {\cal V}_{a}}
{\Gamma_{a,+}T_++\Gamma_{a,-}T_-}
\right].
\end{aligned}
\label{Kramers}
\end{equation}
The prefactor $\Gamma_0$ depends on the local curvatures and the detailed dissipative dynamics. Equation~(\ref{Kramers}) should not be used outside the weak-noise, high-barrier and effectively Markovian regime without the appropriate non-Markovian generalization.

\section{Schwarzschild--Schwarzschild Thin-Shell Wormholes}
\label{sec:SchSch}

We now specialize the geometric sector to two Schwarzschild exteriors. In this section the stochastic and mean fluxes are treated in a test-field/coarse-grained approximation: their backreaction on the bulk metrics is neglected and the parameters $M_\pm$ are held fixed. A finite self-consistent energy flux that appreciably changes the bulk geometry would require a nonstationary exterior, for example a Vaidya-type mass function, and lies outside the present Schwarzschild-background approximation.
\begin{equation}
f_\pm(r)=1-\frac{2M_\pm}{r},
\qquad
\gamma_\pm=
\sqrt{1-\frac{2M_\pm}{a}+\dot a^2}.
\end{equation}
The surface energy density and pressure are
\begin{equation}
\sigma
=
-\frac{\gamma_++\gamma_-}{4\pi a},
\label{sigmaSch}
\end{equation}
and
\begin{equation}
P
=
\frac1{8\pi}
\left[
\frac{\ddot a+M_+/a^2}{\gamma_+}
+
\frac{\ddot a+M_-/a^2}{\gamma_-}
+
\frac{\gamma_++\gamma_-}{a}
\right].
\label{pressureSch}
\end{equation}
The local acceleration scales are
\begin{equation}
\kappa_\pm
=
\frac{\left|\ddot a+M_\pm/a^2\right|}
{\gamma_\pm},
\qquad
T_\pm=\frac{\hbar\kappa_\pm}{2\pi},
\end{equation}
and
\begin{equation}
T_{\rm acc}
=
\frac{\hbar}{4\pi}(\kappa_++\kappa_-).
\end{equation}
The surface-stress scale is
\begin{equation}
\kappa_{\rm stress}
=
\frac12\left|
\frac{\ddot a+M_+/a^2}{\gamma_+}
+
\frac{\ddot a+M_-/a^2}{\gamma_-}
+
\frac{\gamma_++\gamma_-}{a}
\right|.
\end{equation}

\subsection{Symmetric static branch and local stochastic closure}
\label{sec:SchClosure}

We now consider the symmetric configuration $M_+=M_-=M$. Then
\begin{equation}
\gamma=\sqrt{1-\frac{2M}{a}+\dot a^2},
\qquad
\sigma=-\frac{\gamma}{2\pi a},
\label{SchSymSigma}
\end{equation}
and
\begin{equation}
P
=
\frac1{4\pi}
\left[
\frac{\ddot a+M/a^2}{\gamma}
+
\frac{\gamma}{a}
\right].
\label{Psymmetric}
\end{equation}
The corresponding acceleration and surface-stress temperatures are
\begin{equation}
T_{\rm acc}
=
\frac{\hbar}{2\pi}
\frac{\left|\ddot a+M/a^2\right|}{\gamma},
\label{AccelerationTemperatureSym}
\end{equation}
and
\begin{equation}
T_{\rm stress}
=
\frac{\hbar}{2\pi}
\left|
\frac{\ddot a+M/a^2}{\gamma}
+
\frac{\gamma}{a}
\right|.
\label{StressTemperatureSym}
\end{equation}

For a static throat $a=a_0>2M$, define
\begin{equation}
f_0=1-\frac{2M}{a_0}>0.
\end{equation}
The static surface quantities are
\begin{equation}
\sigma_0
=
-\frac{\sqrt{f_0}}{2\pi a_0},
\qquad
P_0
=
\frac{a_0-M}{4\pi a_0^2\sqrt{f_0}}.
\label{SchStaticSigmaP}
\end{equation}
The shell area and internal energy are
\begin{equation}
A_0=4\pi a_0^2,
\qquad
U_0=\sigma_0A_0
=-2a_0\sqrt{f_0}.
\label{SchStaticAU}
\end{equation}
The static temperature scales become
\begin{equation}
T_{\rm acc}^{(0)}
=
\frac{\hbar M}{2\pi a_0^2\sqrt{f_0}},
\qquad
T_{\rm stress}^{(0)}
=
\frac{\hbar(a_0-M)}{2\pi a_0^2\sqrt{f_0}},
\label{SchStaticTemperatures}
\end{equation}
so that
\begin{equation}
\frac{T_{\rm stress}^{(0)}}{T_{\rm acc}^{(0)}}
=
\frac{a_0-M}{M}.
\label{SchTempRatio}
\end{equation}
Thus the two scales are distinct at a generic static radius, but approach one another when $a_0\rightarrow2M^+$.

To close the local mechanical problem without imposing an arbitrary global equation of state, we introduce only the linearized constitutive derivative
\begin{equation}
\beta_0^2
\equiv
\left(\frac{dP}{d\sigma}\right)_0\,,\qquad
\delta P=\beta_0^2\,\delta\sigma+O(\delta\sigma^2).
\label{betaEOS}
\end{equation}
The notation $\beta_0^2$ is conventional and should not be interpreted, in the present exotic-shell setting, as necessarily defining a physical or causal sound speed.

From Eq.~(\ref{SchSymSigma}),
\begin{equation}
\dot a^2+V(a)=0\,,\quad
V(a)
=
1-\frac{2M}{a}
-\left[2\pi a\sigma(a)\right]^2.
\label{SchPotential}
\end{equation}
The conservation equation gives
\begin{equation}
\sigma'(a)
=
-\frac{2}{a}\left[\sigma(a)+P(a)\right].
\label{SchConservationDerivative}
\end{equation}
At the static point, $V(a_0)=V'(a_0)=0$, and use of Eq.~(\ref{betaEOS}) yields
\begin{eqnarray}
V''(a_0)
&=&
-\frac{2}{a_0^3(a_0-2M)}
\Big[
a_0^2-3Ma_0+3M^2
\nonumber\\
&&+
2\beta_0^2(a_0-2M)(a_0-3M)
\Big].
\label{SchVpp}
\end{eqnarray}
Hence
\begin{eqnarray}
	\omega_0^2
	=
	\frac12V''(a_0)
	&=&-\frac{1}{a_0^3(a_0-2M)}
	\Big[
	a_0^2-3Ma_0+3M^2
	\nonumber\\
	&&+
	2\beta_0^2(a_0-2M)(a_0-3M)
	\Big].
	\label{SchOmegaExplicit}
\end{eqnarray}
Linear mechanical stability requires $\omega_0^2>0$, or equivalently
\begin{equation}
a_0^2-3Ma_0+3M^2
+
2\beta_0^2(a_0-2M)(a_0-3M)
<0.
\label{SchStabilityCondition}
\end{equation}

For $2M<a_0<3M$, this condition may be written as
\begin{equation}
\beta_0^2
>
\frac{a_0^2-3Ma_0+3M^2}
{2(a_0-2M)(3M-a_0)}.
\label{SchBetaStability}
\end{equation}
Thus a stable static branch approaching the Schwarzschild radius cannot maintain a finite constitutive slope. Setting $a_0=2M+\epsilon_0$ gives
\begin{equation}
\beta_0^2
>
\frac{M}{2\epsilon_0}
+O(1).
\label{SchBetaNearHorizon}
\end{equation}
This observation is important when comparing the local static closure with the quasi-horizon scaling considered below: the latter is a conditional reduced-model scaling and is not automatically realized by a finite-$\beta_0^2$ static equation of state.

As a useful check, a globally linear choice $P=w\sigma$ tuned to support the same static radius would require
\begin{equation}
w
=
-\frac{a_0-M}{2(a_0-2M)},
\end{equation}
for which
\begin{equation}
\omega_0^2
=
-\frac{M}{a_0^2(a_0-2M)}<0.
\end{equation}
Thus this simplest global linear equation of state does not provide a stable static branch. This illustrates why the local constitutive parametrization (\ref{betaEOS}) is preferable here: it closes the fluctuation analysis without attributing an unjustified global matter model to the shell.

We now include a local radial damping term and write
\begin{equation}
F_{\rm Sch}(a,v)
=
-\frac12V'(a)-\gamma_0v,
\label{SchClosedDrift}
\end{equation}
with $\gamma_0>0$ treated as an effective open-system response coefficient. Around $(a_0,0)$,
\begin{equation}
\dot{\delta a}=\delta v,
\qquad
\dot{\delta v}
=
-\omega_0^2\delta a-\gamma_0\delta v+\eta(\tau).
\label{SchLinearClosed}
\end{equation}

For identical local-equilibrium stochastic environments,
\begin{equation}
\mu_+=\mu_-=\mu_{\rm eff},
\qquad
T_+=T_-=T_{\rm acc}^{(0)},
\end{equation}
and the phase-space diffusion coefficient is
\begin{equation}
D_{v0}
=
2\mu_{\rm eff}T_{\rm acc}^{(0)}
=
\frac{\mu_{\rm eff}\hbar M}
{\pi a_0^2\sqrt{f_0}}.
\label{SchDvExplicit}
\end{equation}
The stationary covariances are therefore
\begin{equation}
\left\langle(\delta v)^2\right\rangle
=
\frac{\mu_{\rm eff}\hbar M}
{\pi\gamma_0a_0^2\sqrt{f_0}}\,,\qquad
\langle\delta a\,\delta v\rangle=0,
\label{SchVelocityExplicit}
\end{equation}
and
\begin{equation}
\left\langle(\delta a)^2\right\rangle
=
\frac{\mu_{\rm eff}\hbar M}
{\pi\gamma_0\omega_0^2a_0^2\sqrt{f_0}}.
\label{SchRadialExplicit}
\end{equation}

\subsection{Overdamped reduction and stationary distribution}

In the overdamped regime $\gamma_0\gg\omega_0$, adiabatic elimination of the velocity gives
\begin{equation}
{\cal A}(a)
=
-\frac{V'(a)}{2\gamma_0},
\qquad
D_a
=
\frac{D_{v0}}{\gamma_0^2}.
\label{SchOverdampedReduction}
\end{equation}
Consequently,
\begin{equation}
\lambda_a
=
-\left(\frac{d{\cal A}}{da}\right)_0
=
\frac{V''(a_0)}{2\gamma_0}
=
\frac{\omega_0^2}{\gamma_0}.
\label{SchLambdaExplicit}
\end{equation}
Identifying
\begin{equation}
\Gamma_{a,\rm eff}
=
\frac{\mu_{\rm eff}}{\gamma_0^2},
\end{equation}
one recovers
\begin{equation}
D_a
=
2\Gamma_{a,\rm eff}T_{\rm acc}^{(0)}.
\label{DiffusionSym}
\end{equation}
The two descriptions are mutually consistent because
\begin{equation}
\frac{D_a}{\lambda_a}
=
\frac{D_{v0}}{\gamma_0\omega_0^2},
\label{VarianceConsistency}
\end{equation}
so the stationary radial variance obtained after adiabatic elimination agrees with the phase-space result.

For approximately constant $\gamma_0$,
\begin{equation}
{\cal V}_a(a)
=
-\int^a{\cal A}(x)\,dx
=
\frac{V(a)}{2\gamma_0}+{\rm const.}
\label{SchQuasipotential}
\end{equation}
Near the stable point,
\begin{equation}
{\cal V}_a(a)
\simeq
{\cal V}_a(a_0)
+
\frac{\lambda_a}{2}(a-a_0)^2,
\end{equation}
and the stationary zero-current density is
\begin{equation}
{\cal P}_{\rm st}(a)
=
\sqrt{\frac{\lambda_a}{2\pi D_a}}
\exp\left[
-\frac{\lambda_a(a-a_0)^2}{2D_a}
\right].
\label{SchStationaryDistribution}
\end{equation}
Equivalently,
\begin{equation}
{\cal P}_{\rm st}(a)
=
\sqrt{\frac{\gamma_0\omega_0^2}{2\pi D_{v0}}}
\exp\left[
-\frac{\gamma_0\omega_0^2}{2D_{v0}}(a-a_0)^2
\right].
\end{equation}

\subsection{Constitutive fluctuations and quasistatic sensitivities}

A distinction is required between two different derivatives. First, the transparent static junction family defines the geometrical functions
\begin{equation}
\sigma_0(a)
=
-\frac{\sqrt{1-2M/a}}{2\pi a},
\qquad
\frac{d\sigma_0}{da}
=
\frac{a-3M}{2\pi a^3\sqrt{1-2M/a}},
\label{SigmaDerivativeSch}
\end{equation}
and
\begin{equation}
P_0^{\rm stat}(a)
=
\frac{a-M}{4\pi a^2\sqrt{1-2M/a}},
\end{equation}
with
\begin{equation}
\frac{dP_0^{\rm stat}}{da}
=
-\frac{a^2-3aM+3M^2}
{4\pi a^4(1-2M/a)^{3/2}}.
\label{PressureDerivativeSch}
\end{equation}
For $2M<a<3M$ their derivative ratio is
\begin{equation}
\frac{dP_0^{\rm stat}/da}{d\sigma_0/da}
=
\frac{a^2-3Ma+3M^2}
{2(a-2M)(3M-a)}
\equiv\beta_{\rm crit}^2(a).
\label{StaticLocusSlope}
\end{equation}
Comparison with Eq.~(\ref{SchBetaStability}) shows that this is precisely the marginal-stability value. Substitution into Eq.~(\ref{SchOmegaExplicit}) gives
\begin{equation}
\omega_0^2\big|_{\beta_0^2=\beta_{\rm crit}^2}=0.
\label{StaticLocusMarginal}
\end{equation}
Thus moving through the one-parameter family of distinct static junction solutions is not the same operation as perturbing one stable shell with a prescribed constitutive slope $\beta_0^2>\beta_{\rm crit}^2$. This observation fixes the stochastic interpretation of the pressure sector.

Within the stable reduced constitutive closure, the surface-density fluctuation is
\begin{equation}
\left\langle(\delta\sigma)^2\right\rangle_0
=
\frac{(a_0-3M)^2}
{4\pi^2a_0^6f_0}
\frac{D_a}{\lambda_a},
\label{ExplicitSigmaVarianceSch}
\end{equation}
while Eq.~(\ref{betaEOS}) requires
\begin{equation}
\delta P=\beta_0^2\,\delta\sigma,
\qquad
\left\langle(\delta P)^2\right\rangle_0
=
(\beta_0^2)^2
\left\langle(\delta\sigma)^2\right\rangle_0.
\label{ExplicitPressureVarianceSch}
\end{equation}
This is the finite-time/coarse-grained pressure variance associated with the chosen constitutive response. The instantaneous pressure of the full white-noise phase-space process remains distribution-valued because it contains $\dot v$, as discussed in Sec.~\ref{sec:phaseSpaceStochastic}.

For the surface-stress scale, $P_0>0$ in the static Schwarzschild branch and therefore, to linear order,
\begin{equation}
\delta T_{\rm stress}
=2\hbar\,\delta P
=2\hbar\beta_0^2\,\delta\sigma,
\end{equation}
so that
\begin{equation}
\left\langle(\delta T_{\rm stress})^2\right\rangle_0
=
4\hbar^2(\beta_0^2)^2
\left\langle(\delta\sigma)^2\right\rangle_0.
\label{ExplicitStressTemperatureVarianceSch}
\end{equation}

The acceleration scale requires a separate qualification. The exact quantity $T_{\rm acc}[a,v,\dot v]$ contains the stochastic acceleration and therefore has no finite instantaneous variance in the ideal white-noise model. A useful reduced diagnostic is instead the \emph{quasistatic static-locus sensitivity}
\begin{equation}
T_{\rm acc}^{\rm qs}(a)
=
\frac{\hbar M}{2\pi a^2\sqrt{1-2M/a}},
\end{equation}
for which
\begin{equation}
\frac{dT_{\rm acc}^{\rm qs}}{da}
=
\frac{\hbar M(3M-2a)}
{2\pi a^4(1-2M/a)^{3/2}}.
\end{equation}
Its induced radial sensitivity is
\begin{equation}
\left\langle(\delta T_{\rm acc}^{\rm qs})^2\right\rangle_0
=
\frac{\hbar^2M^2(3M-2a_0)^2}
{4\pi^2a_0^8f_0^3}
\frac{D_a}{\lambda_a}.
\label{ExplicitTemperatureVarianceSch}
\end{equation}
This quantity characterizes how the static acceleration scale changes along the reduced radial coordinate; it is not the ultraviolet-sensitive variance of the exact stochastic acceleration observable.

For the shell area,
\begin{equation}
\left\langle(\delta A)^2\right\rangle_0
=
64\pi^2a_0^2\frac{D_a}{\lambda_a},
\label{ExplicitAreaVarianceSch}
\end{equation}
while the reference-branch internal energy $U(a)=-2a\sqrt{1-2M/a}$ gives
\begin{equation}
\left\langle(\delta U)^2\right\rangle_0
=
\frac{4(a_0-M)^2}{a_0^2f_0}
\frac{D_a}{\lambda_a}.
\label{ExplicitEnergyVarianceSch}
\end{equation}

More generally, any pair of smooth observables $\mathcal O_i(a)$ and $\mathcal O_j(a)$ that are genuinely defined on the chosen reduced one-parameter closure obey
\begin{equation}
\left\langle
\delta{\cal O}_i(\tau)\delta{\cal O}_j(0)
\right\rangle
=
{\cal O}'_{i0}{\cal O}'_{j0}
\frac{D_a}{\lambda_a}
e^{-\lambda_a|\tau|}.
\label{SchCrossCorrelation}
\end{equation}
For the constitutive pressure and stress temperature one must use $P'_0=\beta_0^2\sigma'_0$ and $T_{{\rm stress},0}'=2\hbar\beta_0^2\sigma'_0$; for the acceleration scale the same formula applies only to the quasistatic diagnostic $T_{\rm acc}^{\rm qs}(a)$ defined above.

\subsection{Markovian and non-Markovian spectra}

For the symmetric Markovian process,
\begin{equation}
S_a^{\rm M}(\omega)
=
\frac{4\Gamma_{a,\rm eff}T_{\rm acc}^{(0)}}
{\lambda_a^2+\omega^2}
=
\frac{2\Gamma_{a,\rm eff}\hbar M}
{\pi a_0^2\sqrt{f_0}\,(\lambda_a^2+\omega^2)}.
\label{SchMarkovSpectrum}
\end{equation}
Any quasistatic observable obeys
\begin{equation}
S_{\cal O}^{\rm M}(\omega)
=
({\cal O}'_0)^2S_a^{\rm M}(\omega),
\label{SchObservableSpectrum}
\end{equation}
which gives the spectra of any observables that belong to the same reduced one-parameter closure. In particular, $P'_0=\beta_0^2\sigma'_0$ and $T_{{\rm stress},0}'=2\hbar\beta_0^2\sigma'_0$ must be used for the constitutive pressure and surface-stress scale, while the derivative formula for $T_{\rm acc}$ refers only to the quasistatic diagnostic $T_{\rm acc}^{\rm qs}(a)$.

For identical memory sectors, let
\begin{equation}
\Lambda_+=\Lambda_-=\Lambda_{\rm eff}\,,\qquad
\Lambda_{\rm tot}=2\Lambda_{\rm eff}.
\end{equation}
Then
\begin{equation}
{\cal M}(t)
=
\frac{2\Lambda_{\rm eff}}{\tau_c}
e^{-t/\tau_c}\Theta(t)\,,\qquad
\widetilde{\cal M}(\omega)
=
\frac{2\Lambda_{\rm eff}}{1-i\omega\tau_c},
\end{equation}
and
\begin{equation}
S_\zeta(\omega)
=
\frac{4\Gamma_{a,\rm eff}T_{\rm acc}^{(0)}}
{1+\omega^2\tau_c^2}.
\end{equation}
The non-Markovian radial spectrum becomes
\begin{equation}
S_a^{\rm NM}(\omega)
=
\frac{4\Gamma_{a,\rm eff}T_{\rm acc}^{(0)}}
{1+\omega^2\tau_c^2}
\left|
\lambda_{a0}-i\omega
+
\frac{2\Lambda_{\rm eff}}
{1-i\omega\tau_c}
\right|^{-2}.
\label{SchNMSpectrum}
\end{equation}
The memory coefficient $\Lambda_{\rm eff}$ and mobility $\Gamma_{a,\rm eff}$ remain independent phenomenological parameters unless a microscopic generalized fluctuation--dissipation relation is derived.

In the symmetric short-memory limit, Eq.~(\ref{MarkovianMatching}) becomes
\begin{equation}
\lambda_a=\lambda_{a0}+2\Lambda_{\rm eff},
\label{SchMarkovianMatching}
\end{equation}
and Eq.~(\ref{SchNMSpectrum}) correspondingly reduces to the Markovian result (\ref{SchMarkovSpectrum}).

\subsection{Entropy production and detailed balance}

For the symmetric local-equilibrium configuration,
\begin{equation}
T_+=T_-=T_{\rm acc}^{(0)},
\end{equation}
and therefore the thermal affinity vanishes,
\begin{equation}
{\cal X}_T
=
\frac1{T_-}-\frac1{T_+}
=0.
\end{equation}
Within the Onsager description this implies
\begin{equation}
J_Q=0,
\qquad
\Pi_Q=0.
\label{SchReservoirEntropy}
\end{equation}
For the symmetric \emph{Markovian} reduced local-equilibrium model, the stationary reduced Gaussian distribution satisfies $J_a=0$, so that
\begin{equation}
\Pi_{\rm cg}=0,
\qquad
\Phi_{S,\rm cg}=0,
\qquad
\frac{dS_{\rm cg}}{d\tau}=0.
\label{SchReducedEntropy}
\end{equation}
These statements refer to the symmetric local-equilibrium reduced model; they do not exclude hidden microscopic currents beyond the coarse-grained radial description. Nor does $J_a=0$ automatically establish detailed balance for the non-Markovian model when memory and colored-noise kernels are parametrized independently. In that case equilibrium in the enlarged stochastic state space requires the corresponding generalized fluctuation--dissipation relation.

\subsection{Kramers transitions}

The local stable closure developed above has the harmonic quasipotential
\begin{equation}
{\cal V}_a
\simeq
{\cal V}_a(a_0)+\frac{\lambda_a}{2}(a-a_0)^2,
\end{equation}
with a single local minimum. Hence no quasipotential barrier $\Delta{\cal V}_a$ is generated by the local harmonic approximation, and no nontrivial Kramers escape rate follows from this closure alone. A genuine Kramers transition requires a specified global equation of state, nonlinear drift or flux law capable of producing two or more stable fixed points separated by a barrier. The general expression (\ref{Kramers}) therefore remains conditional in the present Schwarzschild--Schwarzschild example.

\subsection{Quasi-horizon scaling of the stochastic fluctuations}

Set
\begin{equation}
a_0=2M+\epsilon_0,\qquad
\varepsilon_0\equiv\frac{\epsilon_0}{2M}\ll1.
\end{equation}
Then $f_0\simeq\varepsilon_0$ and
\begin{equation}
T_{\rm acc}^{(0)}
\simeq
\frac{\hbar}{8\pi M}\,\varepsilon_0^{-1/2},
\qquad
D_a
\simeq
\frac{\Gamma_{a,\rm eff}\hbar}{4\pi M}\,\varepsilon_0^{-1/2}.
\label{QuasiHorizonDa}
\end{equation}
If $\Gamma_{a,\rm eff}$ and $\lambda_a$ remain finite and nonzero in this conditional constant-response reduction, then
\begin{equation}
\langle(\delta a)^2\rangle
\propto\varepsilon_0^{-1/2},
\qquad
\langle(\delta A)^2\rangle
\propto\varepsilon_0^{-1/2},
\end{equation}
and
\begin{equation}
\langle(\delta\sigma)^2\rangle
\propto\varepsilon_0^{-3/2},
\qquad
\langle(\delta U)^2\rangle
\propto\varepsilon_0^{-3/2}.
\label{QHSigmaUScaling}
\end{equation}
The quasistatic acceleration sensitivity defined in Eq.~(\ref{ExplicitTemperatureVarianceSch}) scales as
\begin{equation}
\left\langle(\delta T_{\rm acc}^{\rm qs})^2\right\rangle
\propto\varepsilon_0^{-7/2}.
\label{QHAccSensitivityScaling}
\end{equation}

The pressure and surface-stress sectors depend explicitly on the constitutive slope:
\begin{equation}
\begin{aligned}
\langle(\delta P)^2\rangle
&=(\beta_0^2)^2\langle(\delta\sigma)^2\rangle,\\
\langle(\delta T_{\rm stress})^2\rangle
&=4\hbar^2(\beta_0^2)^2
\langle(\delta\sigma)^2\rangle.
\end{aligned}
\label{QHConstitutiveScaling}
\end{equation}
Thus no universal power can be assigned to these two variances without specifying how $\beta_0^2$ scales. A formally finite $\beta_0^2$ would give the same $\varepsilon_0^{-3/2}$ power as the surface-density variance, but Eq.~(\ref{SchBetaNearHorizon}) shows that such a sequence cannot remain mechanically stable as $a_0\to2M^+$. The minimal stable scaling $\beta_0^2=O(\varepsilon_0^{-1})$ would instead produce an $O(\varepsilon_0^{-7/2})$ constitutive amplification \emph{if} $\lambda_a$ were held fixed. This last proviso is essential because Eq.~(\ref{SchLambdaExplicit}) makes $\lambda_a$ itself dependent on the simultaneous scaling of $\beta_0^2$ and $\gamma_0$. Consequently, none of these powers should be interpreted as a universal stable-horizon limit.

The increasing fluctuations are instead a diagnostic of the breakdown of the local stationary overdamped approximation. In particular, once $\sqrt{\langle(\delta a)^2\rangle}\sim\epsilon_0$, the Gaussian expansion about a fixed $a_0$ is no longer self-consistent, and the full phase-space dynamics, multiplicative noise, non-Markovian response, and gravitational backreaction must be restored before the formal horizon is reached.

\subsection{Asymmetric configurations}

For $M_+\neq M_-$, the kinematical acceleration scales $T_+$ and $T_-$ generally differ. This asymmetry changes the junction stresses and, under the local-thermal/detailed-balance assumption, the stochastic noise strength in Eq.~(\ref{FDTcorrected}). It does not by itself prove a stationary heat current: such a current requires a specified nonequilibrium quantum state and coupling satisfying the assumptions of Sec.~\ref{sec:nonEq}. Accordingly, the asymmetric wormhole may be treated as a gravitational analogue of an interface between two reservoirs only within that effective open-system approximation.

\section{Quasi-Horizon Thermodynamics and Semiclassical Temperature Unification}
\label{sec:quasi}

For the symmetric Schwarzschild--Schwarzschild wormhole, write
\begin{equation}
a(\tau)=2M+\epsilon(\tau)\,,\qquad
0<\epsilon\ll2M.
\end{equation}

Here $\epsilon=a-2M$ is the Schwarzschild radial-coordinate separation from the horizon radius, not the proper radial distance. It is useful to introduce the dimensionless quasi-horizon parameter
\begin{equation}
\varepsilon(\tau)\equiv\frac{\epsilon(\tau)}{2M}\ll1.
\label{dimensionlessQH}
\end{equation}
The proper radial distance is then
\begin{equation}
\ell(\epsilon)
=
\int_{2M}^{2M+\epsilon}
\frac{dr}{\sqrt{1-2M/r}}
=
4M\sqrt{\varepsilon}
+
{\cal O}\!\left(M\varepsilon^{3/2}\right).
\label{properdistance}
\end{equation}

The coordinate limit $\epsilon\to0^+$ alone does not imply a quasi-static approach. From the exact acceleration scale, a trajectory with $\dot\epsilon\to v_0\neq0$ has a finite limiting denominator and need not exhibit a divergent local acceleration temperature. We therefore distinguish this general near-horizon limit from the quasi-static sector and now impose
\begin{equation}
\dot\epsilon^2=O(\varepsilon),
\label{scaling}
\end{equation}
under which $\dot\epsilon\to0$ and
\begin{equation}
f(a)
=
\varepsilon+O(\varepsilon^2)\,,\qquad
\gamma
=
\sqrt{\dot\epsilon^2+\varepsilon}
+O(\varepsilon^{3/2})\,.
\end{equation}
The surface energy density is therefore
\begin{equation}
\sigma
=
-\frac1{4\pi M}
\sqrt{\dot\epsilon^2+\varepsilon}
+
O\!\left(\frac{\varepsilon^{3/2}}{M}\right).
\label{SigmaNear}
\end{equation}

Using
\begin{equation}
\frac{M}{a^2}
=
\frac{1}{4M}+O\!\left(\frac{\varepsilon}{M}\right),
\end{equation}
the pressure may be written consistently to the required order as
\begin{equation}
P
=
\frac1{4\pi}
\frac{\ddot\epsilon+\dfrac1{4M}+O(\varepsilon/M)}
{\sqrt{\dot\epsilon^2+\varepsilon}}
+
O\!\left(\frac{\sqrt{\varepsilon}}{M}\right).
\label{PressureExpansionQH}
\end{equation}
If the leading numerator remains nonzero, the first term dominates and
\begin{equation}
P
\simeq
\frac1{4\pi}
\frac{\ddot\epsilon+1/(4M)}
{\sqrt{\dot\epsilon^2+\varepsilon}}.
\label{PressureNear}
\end{equation}
The acceleration scale has the same leading magnitude,
\begin{equation}
T_{\rm acc}
\simeq
\frac{\hbar}{2\pi}
\frac{\left|\ddot\epsilon+1/(4M)\right|}
{\sqrt{\dot\epsilon^2+\varepsilon}}.
\label{NearTemperature}
\end{equation}
Within the quasi-static scaling (\ref{scaling}), the denominator necessarily tends to zero. A divergent quasi-horizon acceleration scale therefore occurs when the leading numerator does not vanish sufficiently rapidly; the coordinate limit $a\to2M^+$ by itself is not enough without the quasi-static assumption. Equivalently, in the notation used below the relevant denominator condition is
\begin{equation}
\dot\epsilon^2+\varepsilon\to0.
\label{divergencecondition}
\end{equation}

Writing $X=\ddot a+M/a^2$, Eq.~(\ref{Psymmetric}) gives an exact useful bound,
\begin{equation}
\left|T_{\rm stress}-T_{\rm acc}\right|
\leq
\frac{\hbar}{2\pi}\frac{\gamma}{a}.
\label{NearTemperatureBound}
\end{equation}
Hence
\begin{equation}
T_{\rm stress}=T_{\rm acc}+O\!\left(\frac{\hbar\gamma}{a}\right),
\label{NearIdentity}
\end{equation}
and, under the quasi-static scaling $\gamma=O(\sqrt{\varepsilon})$,
\begin{equation}
T_{\rm stress}=T_{\rm acc}+O\!\left(\frac{\hbar\sqrt{\varepsilon}}{M}\right).
\end{equation}
For the two quantities to share the same \emph{leading} scale one further requires $|X|/\gamma\gg\gamma/a$, equivalently $|X|\gg\gamma^2/a$. This condition is automatically satisfied on the strictly static branch, and more generally along quasistatic trajectories for which $X$ approaches a finite nonzero limit; quasistaticity by itself does not exclude $X\to0$. The exact bound (\ref{NearTemperatureBound}) is the robust junction-condition statement.

Within the symmetric reduced radial local-equilibrium ansatz,
\begin{equation}
D_a
=
2\Gamma_{a,\rm eff}T_{\rm acc}.
\label{NearDiffusion}
\end{equation}
Hence $D_a$ diverges only when the condition (\ref{divergencecondition}) makes
$T_{\rm acc}$ diverge.  In that regime the radial stationary distribution
${\cal P}_{\rm st}\propto\exp(-{\cal V}_a/D_a)$ becomes increasingly broad.
The explicit Schwarzschild--Schwarzschild variances derived in
Sec.~\ref{sec:SchSch} show that the radial and surface-density fluctuations,
together with the quasistatic acceleration sensitivity, are strongly amplified
in the near-horizon regime, while the constitutive pressure and surface-stress
sectors depend additionally on the scaling of $\beta_0^2$. This behavior signals the
breakdown of the local stationary overdamped approximation and the need to
restore the full phase-space dynamics, multiplicative noise, backreaction and/or
non-Markovian effects before the formal horizon is reached.

\subsection{Semiclassical particle creation and microscopic stochastic noise}
\label{sec:particle}

The stochastic force introduced above is phenomenological. A possible microscopic contribution is supplied by quantum stress-tensor fluctuations. When the global quantum state is approximately factorized or effectively decohered with respect to the two asymptotic sectors, the associated flux fluctuations may be treated as separate reservoir contributions. More generally the state may be entangled across the wormhole geometry, in which case cross-correlations between the two side fluxes must be retained and the independent-reservoir reduction is not valid.

For a time-dependent geometry, positive-frequency modes in the asymptotic past may mix with negative-frequency modes in the future through Bogoliubov coefficients, as in the standard curved-spacetime particle-creation framework \cite{BirrellDavies1982,Wald1994}. The corresponding particle number on each side is schematically
\begin{equation}
N_\omega^{(\pm)}
=
\sum_{\omega'}
|\beta_{\omega\omega'}^{(\pm)}|^2.
\end{equation}
For a physical collapse-type exterior one usually has a single asymptotic region, whereas the present wormhole construction possesses two asymptotic ends and can therefore support two distinct asymptotic spectra.

The geometrical origin of approximately thermal particle creation can be encoded in ray-tracing maps
\begin{equation}
U_\pm=p_\pm(u_\pm),
\end{equation}
with peeling functions
\begin{equation}
\kappa_\pm^{\rm peel}(u_\pm)
=
-
\frac{p_\pm''(u_\pm)}
{p_\pm'(u_\pm)}.
\label{peeling}
\end{equation}
When $\kappa_\pm^{\rm peel}$ varies adiabatically on its own time scale, an approximately Planckian spectrum may be associated locally with
\begin{equation}
T_\pm^{\rm peel}
=
\frac{\hbar}{2\pi}
\left|\kappa_\pm^{\rm peel}\right|.
\label{sidepeelT}
\end{equation}
The adiabaticity condition is most naturally stated with respect to the asymptotic retarded time used in the ray-tracing map,
\begin{equation}
\left|
\frac{1}{(\kappa_\pm^{\rm peel})^2}
\frac{d\kappa_\pm^{\rm peel}}{du_\pm}
\right|\ll1.
\label{adiabaticpeel}
\end{equation}

In an asymmetric configuration, the sum of two Planckian spectra with different temperatures is not itself Planckian. If a single positive comparison scale is nevertheless useful, one may define
\begin{equation}
\kappa_{\rm peel}^{\rm eff}
=
\frac12
\left(
\left|\kappa_+^{\rm peel}\right|
+
\left|\kappa_-^{\rm peel}\right|
\right).
\end{equation}
This is only an averaged positive peeling scale; it is not the unique physical temperature of the combined radiation field. In the symmetric case the distinction disappears. When the two side peeling magnitudes coincide, $|\kappa_+^{\rm peel}|=|\kappa_-^{\rm peel}|\equiv|\kappa^{\rm peel}|$, we denote their common temperature scale by
\begin{equation}
T_{\rm peel}\equiv\frac{\hbar}{2\pi}|\kappa^{\rm peel}|.
\label{TpeelDef}
\end{equation}

The quantum stress tensor may be written
\begin{equation}
\hat T_{\mu\nu}
=
\langle\hat T_{\mu\nu}\rangle
+
\hat t_{\mu\nu}\,,\qquad
\langle\hat t_{\mu\nu}\rangle=0.
\end{equation}
Projecting $\hat t_{\mu\nu}$ onto the shell world volume produces stochastic flux fluctuations. In a coarse-grained treatment these can contribute to the effective phase-space forcing $\eta(\tau)$ and, after overdamped reduction, to the reduced forcing $\zeta(\tau)$. A microscopic stochastic-gravity calculation would determine the corresponding noise kernel from the symmetrized stress-tensor correlator rather than assuming it a priori \cite{HuVerdaguer2008,SinhaRavalHu2003,PhillipsHu2001}.

A crucial distinction must be maintained between acceleration temperature and particle-creation temperature. A static shell at $a=a_0>2M$ has nonzero proper acceleration and therefore a nonzero Unruh-like $T_{\rm acc}$ for comoving observers, but the geometry is stationary and does not, merely by virtue of that acceleration, generate asymptotic particle creation. Consequently,
\begin{equation}
T_{\rm peel}=T_{\rm acc}
\end{equation}
is not a generic identity. It can hold only for special dynamical ray-tracing maps satisfying an additional relation between the peeling rate and the shell acceleration. This condition is stated explicitly in the next section.

\subsection{Conditional unification of effective temperatures}
\label{sec:unification}

The quasi-horizon relation between the two local shell scales follows directly from the junction conditions. In the symmetric case,
\begin{equation}
X(\tau)\equiv\ddot a+\frac{M}{a^2},
\qquad
\kappa_{\rm acc}=\frac{|X|}{\gamma},
\qquad
\kappa_{\rm stress}=\left|\frac{X}{\gamma}+\frac{\gamma}{a}\right|.
\end{equation}
The reverse triangle inequality then reproduces the exact bound (\ref{NearTemperatureBound}) and hence
\begin{equation}
T_{\rm stress}=T_{\rm acc}+O\!\left(\frac{\hbar\gamma}{a}\right).
\label{AccStressIdentity}
\end{equation}
Under the quasi-static scaling of Sec.~\ref{sec:quasi}, $\gamma=O(\sqrt{\varepsilon})$. Equality of the \emph{leading} scales additionally requires $|X|\gg\gamma^2/a$, as already stated below Eq.~(\ref{NearIdentity}); this is stronger than the mere coordinate limit $a\to2M^+$.

A peeling temperature joins this equality only under an additional dynamical hypothesis. Suppose that the ray-tracing function is asymptotically exponential,
\begin{equation}
p_\pm(u_\pm)=U_{0\pm}-A_\pm e^{-\alpha_\pm u_\pm},
\end{equation}
so that asymptotically $\kappa_\pm^{\rm peel}=\alpha_\pm$. The peeling scale is a function of the asymptotic null coordinate $u_\pm$, whereas the shell acceleration scale is naturally a function of the proper time $\tau$. Their comparison must therefore be made along the shell trajectory through the composition $u_\pm=u_\pm(\tau)$. Suppose further that the generalized Unruh/peeling condition of Ref.~\cite{Lobo:2026vrn} holds asymptotically in the form
\begin{equation}
\kappa_\pm^{\rm peel}\!\left[u_\pm(\tau)\right]
\simeq
\frac{\left|\ddot a+M/a^2\right|}
{\sqrt{1-2M/a+\dot a^2}},
\label{peelingcondition}
\end{equation}
with the retarded times normalized in the corresponding asymptotic regions. For an exactly exponential asymptotic map this states that $\alpha_\pm$ matches the limiting value of the right-hand side. Together with the adiabaticity condition (\ref{adiabaticpeel}), this implies in the symmetric case
\begin{equation}
T_{\rm peel}\simeq T_{\rm acc}.
\label{PeelTemperatureCorrect}
\end{equation}
The implication is conditional: Eq.~(\ref{peelingcondition}) is not a consequence of $a\to2M$ alone and must be established from the actual null ray-tracing map, its normalization at infinity, and the relation between $u_\pm$ and the shell proper time.

Combining the robust junction-condition result with the additional peeling hypothesis gives

\begin{equation}
T_{\rm stress}=T_{\rm acc}+O\!\left(\frac{\hbar\gamma}{a}\right)
\label{AccStressNearIdentity}
\end{equation}

and, provided Eqs.~(\ref{peelingcondition}) and (\ref{adiabaticpeel}) hold,
\begin{equation}
T_{\rm peel}\simeq T_{\rm acc}.
\label{ConditionalPeelIdentity}
\end{equation}
Within the symmetric local-equilibrium stochastic ansatz this further gives
\begin{equation}
D_a
=
2\Gamma_{a,\rm eff}T_{\rm acc}
\simeq
2\Gamma_{a,\rm eff}T_{\rm stress},
\label{UnifiedNoiseScale1}
\end{equation}
and, conditionally,

\begin{equation}
D_a
\simeq
2\Gamma_{a,\rm eff}T_{\rm peel}
\label{UnifiedNoiseScale}
\end{equation}
when the peeling condition is satisfied. The approximate sign is essential because the peeling relation itself is asymptotic rather than an exact identity.

This formulation separates three logically distinct statements. First, the acceleration scale is kinematical and observer dependent. Second, the surface-stress scale follows from the Israel stress and approaches the acceleration scale geometrically near the quasi-horizon. Third, the peeling scale is semiclassical and depends on the global null ray-tracing map. Their coincidence is therefore a conditional dynamical correspondence, not a universal consequence of gravitational blueshift.

The Tolman relation \cite{Tolman1930,Tolman1934} provides a useful comparison but should not be identified with Eq.~(\ref{peelingcondition}). For a static Schwarzschild field,
\begin{equation}
T_{\rm local}(r)
=
\frac{T_\infty}{\sqrt{1-2M/r}},
\end{equation}
so a local temperature may diverge while the temperature normalized at infinity remains finite. Likewise, a local acceleration scale that diverges near $2M$ does not by itself imply a divergent asymptotic particle spectrum. The actual asymptotic spectrum must be determined from the ray-tracing function and its normalization at infinity.

\section{Conclusions}
\label{sec:conclusion}

We have developed a controlled effective framework for the stochastic dynamics and semiclassical thermodynamics of dynamical thin-shell wormholes. The central organizational principle is to keep separate the pieces fixed by the junction geometry from those introduced through constitutive physics and open-system coarse graining. The radius and radial velocity $(a,\dot a)$ form the fundamental stochastic phase space, whereas $\sigma$, $P$, and the local temperature scales are induced observables. A one-dimensional Langevin--Fokker--Planck model in $a$ is therefore justified only after an overdamped, adiabatic, or quasistatic reduction.

A second essential distinction concerns conservation. The standard thin-shell potential and its stability frequency arise from the transparent conservative reference shell. Dissipation, mean environmental forces, and stochastic forcing are then added as open-system corrections to the radial drift. Once these terms are present, the conservative first integral is not simultaneously imposed as an exact constraint on every noisy trajectory. This makes the local mechanical stiffness and the phenomenological response coefficients logically compatible rather than competing descriptions of the same dynamics.

For the symmetric Schwarzschild--Schwarzschild throat, the conservative reference problem gives the explicit stability frequency in terms of $M$, $a_0$, and the local constitutive slope $\beta_0^2=(dP/d\sigma)_0$. In the interval $2M<a_0<3M$, stability requires a lower bound on $\beta_0^2$ that diverges as $a_0\to2M^+$. The simplest globally linear equation of state $P=w\sigma$ adjusted to support the same static radius remains unstable. We also identified an important geometric fact: the ratio of the derivatives obtained by moving through the family of static junction solutions, $(dP_0^{\rm stat}/da)/(d\sigma_0/da)$, is exactly the marginal-stability value $\beta_{\rm crit}^2$. Consequently, that static-locus derivative cannot represent the constitutive pressure response of a stable stochastic shell.

This distinction fixes the fluctuation sector. The radial and surface-density variances follow directly from the reduced Ornstein--Uhlenbeck process, while the pressure fluctuation must obey $\delta P=\beta_0^2\delta\sigma$ and the surface-stress fluctuation follows from the same constitutive response. By contrast, the exact acceleration temperature depends on the stochastic acceleration and has an ultraviolet-sensitive instantaneous variance in the white-noise idealization. The derivative of the static acceleration scale is therefore retained only as a quasistatic radial sensitivity. The same separation is required when constructing cross-correlations and observable spectra.

The Markovian and non-Markovian sectors are otherwise mutually consistent. The overdamped relations $\lambda_a=\omega_0^2/\gamma_0$ and $D_a=D_{v0}/\gamma_0^2$ reproduce the phase-space stationary radial variance. In the generalized Langevin description, colored forcing modifies the noise spectrum while the memory kernel modifies the susceptibility. Their amplitudes are kept independent at the phenomenological level; a genuine generalized fluctuation--dissipation theorem would have to derive both from the same microscopic reservoir correlation functions and retarded response kernel. Likewise, the Kramers escape formula is conditional on a global nonlinear drift with distinct stable or metastable states; the local harmonic Schwarzschild closure alone contains no finite escape barrier.

The quasi-horizon analysis exhibits both a robust result and a limitation. From the exact junction pressure one obtains $|T_{\rm stress}-T_{\rm acc}|\leq\hbar\gamma/(2\pi a)$, so the two local scales approach one another additively as $\gamma\to0$ and share the same leading behavior whenever the acceleration numerator dominates over $\gamma^2/a$. At the same time, the reduced stochastic variances become increasingly sensitive to the approach toward $2M$. Their detailed powers are not universal: in particular, the pressure and stress-temperature variances depend explicitly on the scaling of $\beta_0^2$, while mechanical stability itself forces this slope to grow near the horizon. The apparent amplification should therefore be viewed primarily as a diagnostic for the breakdown of the constant-coefficient stationary overdamped approximation.

Finally, the semiclassical peeling scale remains conceptually distinct from both local shell scales. A static accelerated shell can have $T_{\rm acc}\neq0$ without producing an asymptotic particle flux. The relation $T_{\rm peel}\simeq T_{\rm acc}$ requires the actual null ray-tracing map to satisfy the additional generalized peeling condition, together with the relevant adiabaticity and asymptotic normalization requirements. Only under those additional assumptions does one obtain the conditional correspondence $T_{\rm stress}\simeq T_{\rm acc}\simeq T_{\rm peel}$.

The framework should therefore be interpreted as a self-consistent local effective theory, not as a microscopic model of the exotic shell matter. The natural next steps are to supply a physically motivated global equation of state or matter action, derive the dissipative and stochastic kernels from a specified quantum state and shell--field coupling, and include self-consistent backreaction when the exchanged energy is large enough that fixed Schwarzschild masses are no longer adequate. These developments would determine the actual domain of validity of the Markovian, local-equilibrium, and quasistatic approximations and would decide whether genuine noise-induced transitions occur in a fully specified thin-shell model.

\acknowledgments{
MER thanks Conselho Nacional de Desenvolvimento Cient\'ifico e Tecnol\'ogico - CNPq, Brazil, for partial financial support. FSNL acknowledges support from the Funda\c{c}\~{a}o para a Ci\^encia e a Tecnologia (FCT) Scientific Employment Stimulus contract with reference CEECINST/00032/2018, and funding through the research grant UID/04434/2025.}



\begin{thebibliography}{66}

\bibitem{Ashtekar2004}
Abhay Ashtekar and Badri Krishnan.
\newblock Isolated and dynamical horizons and their applications.
\newblock {\em Living Rev. Relativity}, 7:10, 2004.

\bibitem{Barcelo2006}
Carlos Barceló, Stefano Liberati, Sebastiano Sonego, and Matt Visser.
\newblock Hawking-like radiation does not require a trapped region.
\newblock {\em Phys. Rev. Lett.}, 97:171301, 2006.

\bibitem{Barcelo2011}
Carlos Barceló, Stefano Liberati, Sebastiano Sonego, and Matt Visser.
\newblock Minimal conditions for the existence of a hawking-like flux.
\newblock {\em Phys. Rev. D}, 83:041501, 2011.

\bibitem{Bardeen1973}
James~M. Bardeen, Brandon Carter, and Stephen~W. Hawking.
\newblock The four laws of black hole mechanics.
\newblock {\em Commun. Math. Phys.}, 31:161--170, 1973.

\bibitem{Bekenstein1972}
Jacob~D. Bekenstein.
\newblock Black holes and the second law.
\newblock {\em Lett. Nuovo Cim.}, 4:737--740, 1972.

\bibitem{Bekenstein1973}
Jacob~D. Bekenstein.
\newblock Black holes and entropy.
\newblock {\em Phys. Rev. D}, 7:2333--2346, 1973.

\bibitem{BirrellDavies1982}
N.~D. Birrell and P.~C.~W. Davies.
\newblock {\em Quantum Fields in Curved Space}.
\newblock Cambridge University Press, 1982.

\bibitem{Booth2005}
Ivan Booth.
\newblock Black hole boundaries.
\newblock {\em Can. J. Phys.}, 83:1073--1099, 2005.

\bibitem{CalzettaHu2008}
Esteban~A. Calzetta and B.~L. Hu.
\newblock {\em Nonequilibrium Quantum Field Theory}.
\newblock Cambridge University Press, 2008.

\bibitem{Darmois1927}
Georges Darmois.
\newblock Les Équations de la gravitation einsteinienne.
\newblock {\em Mémorial des Sciences Mathématiques}, 25, 1927.

\bibitem{Einstein1905}
Albert Einstein.
\newblock Über die von der molekularkinetischen theorie der wärme geforderte
  bewegung von in ruhenden flüssigkeiten suspendierten teilchen.
\newblock {\em Annalen der Physik}, 322(8):549--560, 1905.

\bibitem{Einstein1935}
Albert Einstein and Nathan Rosen.
\newblock The particle problem in the general theory of relativity.
\newblock {\em Phys. Rev.}, 48:73--77, 1935.

\bibitem{Faraoni2015}
Valerio Faraoni.
\newblock {\em Cosmological and Black Hole Apparent Horizons}.
\newblock Springer, 2015.

\bibitem{Fokker1914}
Adriaan~D. Fokker.
\newblock Die mittlere energie rotierender elektrischer dipole im
  strahlungsfeld.
\newblock {\em Annalen der Physik}, 348:810--820, 1914.

\bibitem{Gardiner2009}
Crispin~W. Gardiner.
\newblock {\em Stochastic Methods}.
\newblock Springer, 4 edition, 2009.

\bibitem{Gibbons1977}
G.~W. Gibbons and Stephen~W. Hawking.
\newblock Action integrals and partition functions in quantum gravity.
\newblock {\em Phys. Rev. D}, 15:2752--2756, 1977.

\bibitem{Hawking1974}
Stephen~W. Hawking.
\newblock Black hole explosions?
\newblock {\em Nature}, 248:30--31, 1974.

\bibitem{Hawking1975}
Stephen~W. Hawking.
\newblock Particle creation by black holes.
\newblock {\em Commun. Math. Phys.}, 43:199--220, 1975.

\bibitem{Hayward1999}
Sean~A. Hayward.
\newblock Dynamic black-hole entropy.
\newblock {\em Phys. Rev. Lett.}, 81:4557--4559, 1998.

\bibitem{Hayward1998}
Sean~A. Hayward.
\newblock Unified first law of black-hole dynamics and relativistic
  thermodynamics.
\newblock {\em Class. Quantum Grav.}, 15:3147--3162, 1998.

\bibitem{HuVerdaguer2008}
B.~L. Hu and Enric Verdaguer.
\newblock Stochastic gravity: Theory and applications.
\newblock {\em Living Reviews in Relativity}, 11:3, 2008.

\bibitem{Israel1966}
Werner Israel.
\newblock Singular hypersurfaces and thin shells in general relativity.
\newblock {\em Nuovo Cimento B}, 44:1--14, 1966.

\bibitem{Jacobson1995}
Ted Jacobson.
\newblock Thermodynamics of spacetime: The einstein equation of state.
\newblock {\em Phys. Rev. Lett.}, 75:1260--1263, 1995.

\bibitem{Kramers1940}
Hendrik~A. Kramers.
\newblock Brownian motion in a field of force and the diffusion model of
  chemical reactions.
\newblock {\em Physica}, 7:284--304, 1940.

\bibitem{Kubo1966}
Ryogo Kubo.
\newblock The fluctuation-dissipation theorem.
\newblock {\em Reports on Progress in Physics}, 29:255--284, 1966.

\bibitem{Langevin1908}
Paul Langevin.
\newblock Sur la théorie du mouvement brownien.
\newblock {\em Comptes Rendus}, 146:530--533, 1908.

\bibitem{Lobo2017}
Francisco S.~N. Lobo, editor.
\newblock {\em Wormholes, Warp Drives and Energy Conditions}.
\newblock Springer, 2017.

\bibitem{Lobo:2026vrn}
Francisco S.~N. Lobo and Manuel~E. Rodrigues.
\newblock {Thermodynamics of thin-shell wormholes}.
\newblock 6 2026.

\bibitem{Morris1988}
Michael~S. Morris and Kip~S. Thorne.
\newblock Wormholes in spacetime and their use for interstellar travel.
\newblock {\em Am. J. Phys.}, 56:395--412, 1988.

\bibitem{Morris1988b}
Michael~S. Morris, Kip~S. Thorne, and Ulvi Yurtsever.
\newblock Wormholes, time machines and the weak energy condition.
\newblock {\em Phys. Rev. Lett.}, 61:1446--1449, 1988.

\bibitem{Nielsen2009}
Alex~B. Nielsen.
\newblock Black holes without event horizons.
\newblock {\em Gen. Relativ. Gravit.}, 41:1539--1584, 2009.

\bibitem{Onsager1931a}
Lars Onsager.
\newblock Reciprocal relations in irreversible processes. i.
\newblock {\em Physical Review}, 37:405--426, 1931.

\bibitem{Onsager1931b}
Lars Onsager.
\newblock Reciprocal relations in irreversible processes. ii.
\newblock {\em Physical Review}, 38:2265--2279, 1931.

\bibitem{Padmanabhan2010}
T.~Padmanabhan.
\newblock Thermodynamical aspects of gravity: New insights.
\newblock {\em Rep. Prog. Phys.}, 73:046901, 2010.

\bibitem{PhillipsHu2001}
N.~G. Phillips and B.~L. Hu.
\newblock Noise kernel in stochastic gravity.
\newblock {\em Physical Review D}, 63:104001, 2001.

\bibitem{Planck1917}
Max Planck.
\newblock Über einen satz der statistischen dynamik und seine erweiterung in
  der quantentheorie.
\newblock {\em Sitzungsberichte der Preussischen Akademie der Wissenschaften},
  pages 324--341, 1917.

\bibitem{PoissonVisser1995}
Eric Poisson and Matt Visser.
\newblock Thin-shell wormholes: Linearization stability.
\newblock {\em Phys. Rev. D}, 52:7318--7321, 1995.

\bibitem{Risken1989}
Hannes Risken.
\newblock {\em The Fokker--Planck Equation}.
\newblock Springer, 2 edition, 1989.

\bibitem{Seifert2005}
Udo Seifert.
\newblock Entropy production along a stochastic trajectory and an integral
  fluctuation theorem.
\newblock {\em Physical Review Letters}, 95:040602, 2005.

\bibitem{Seifert2012}
Udo Seifert.
\newblock Stochastic thermodynamics, fluctuation theorems and molecular
  machines.
\newblock {\em Reports on Progress in Physics}, 75:126001, 2012.

\bibitem{Sekimoto2010}
Ken Sekimoto.
\newblock {\em Stochastic Energetics}.
\newblock Springer, 2 edition, 2010.

\bibitem{Thorne1986}
Kip~S. Thorne, Richard~H. Price, and Douglas~A. Macdonald, editors.
\newblock {\em Black Holes: The Membrane Paradigm}.
\newblock Yale University Press, New Haven, 1986.

\bibitem{Tolman1930}
R.~C. Tolman.
\newblock On the weight of heat and thermal equilibrium in general relativity.
\newblock {\em Phys. Rev.}, 35:904--924, 1930.

\bibitem{Tolman1934}
R.~C. Tolman.
\newblock {\em Relativity, Thermodynamics and Cosmology}.
\newblock Oxford University Press, Oxford, 1934.

\bibitem{Unruh1976}
William~G. Unruh.
\newblock Notes on black hole evaporation.
\newblock {\em Phys. Rev. D}, 14:870--892, 1976.

\bibitem{Verlinde2011}
Erik Verlinde.
\newblock On the origin of gravity and the laws of newton.
\newblock {\em JHEP}, 04:029, 2011.

\bibitem{Visser1989}
Matt Visser.
\newblock Traversable wormholes from surgically modified schwarzschild
  spacetimes.
\newblock {\em Nucl. Phys. B}, 328:203--212, 1989.

\bibitem{VisserBook}
Matt Visser.
\newblock {\em Lorentzian Wormholes: From Einstein to Hawking}.
\newblock AIP Press, New York, 1995.

\bibitem{Wald1994}
R.~M. Wald.
\newblock {\em Quantum Field Theory in Curved Spacetime and Black Hole
  Thermodynamics}.
\newblock University of Chicago Press, Chicago, 1994.

\bibitem{Zwanzig1961}
Robert Zwanzig.
\newblock Memory effects in irreversible thermodynamics.
\newblock {\em Physical Review}, 124:983--992, 1961.

\bibitem{Garcia2012}

N. M. Garc\'ia, F. S. N. Lobo, and M. Visser, Phys. Rev. D \textbf{86}, 044026 (2012), arXiv:1112.2057, doi:10.1103/PhysRevD.86.044026.

\bibitem{Eiroa2008}

E. F. Eiroa, Phys. Rev. D \textbf{78}, 024018 (2008), arXiv:0805.1403.

\bibitem{LoboCrawford2005}

F. S. N. Lobo and P. Crawford, Class. Quantum Grav. \textbf{22}, 4869 (2005), arXiv:gr-qc/0507063.

\bibitem{Rippentrop2025}
T. S. Rippentrop, A. Bera, and M. Ishak, Phys. Rev. D \textbf{112}, 124069 (2025), arXiv:2507.00315, doi:10.1103/gpjg-vgwl.

\bibitem{EiroaFigueroa2016}

E. F. Eiroa and G. Figueroa-Aguirre, Eur. Phys. J. C \textbf{76}, 132 (2016), arXiv:1511.02806.

\bibitem{Martinez1996}

E. A. Martinez, Phys. Rev. D \textbf{53}, 7062 (1996), arXiv:gr-qc/9601037.

\bibitem{Forghani2019}

S. D. Forghani, S. H. Mazharimousavi, and M. Halilsoy, Int. J. Mod. Phys. D \textbf{28}, 1950142 (2019), arXiv:1812.04340, doi:10.1142/S0218271819501426.

\bibitem{Eiroa2024}

E. F. Eiroa, G. Figueroa-Aguirre, M. L. Pe\~nafiel, and S. E. Perez Bergliaffa, Eur. Phys. J. C \textbf{84}, 1160 (2024), arXiv:2408.14328.

\bibitem{Reboucas2026}
J. A. Rebou\c{c}as, E. Otoniel, and F. S. N. Lobo, ``Thin-shell wormholes in cosmic voids,'' arXiv:2607.11995 [gr-qc] (2026).

\bibitem{Barcelo2011JHEP}
C. Barcel\'o, S. Liberati, S. Sonego, and M. Visser, JHEP \textbf{02}, 003 (2011), arXiv:1011.5911.

\bibitem{Visser2003}

M. Visser, Int. J. Mod. Phys. D \textbf{12}, 649 (2003), arXiv:hep-th/0106111.

\bibitem{VanKampen2007}

N. G. van Kampen, \textit{Stochastic Processes in Physics and Chemistry}, 3rd ed. (Elsevier, Amsterdam, 2007).

\bibitem{LiZhangWang2021}

R. Li, K. Zhang, and J. Wang, Phys. Rev. D \textbf{104}, 084076 (2021), arXiv:2102.09439.

\bibitem{WangEtAl2026}
C. Wang, C. Ma, M.-C. He, and B. Wu, Eur. Phys. J. C \textbf{86}, 628 (2026), arXiv:2604.05785, doi:10.1140/epjc/s10052-026-15855-1.

\bibitem{SinhaRavalHu2003}

S. Sinha, A. Raval, and B. L. Hu, Found. Phys. \textbf{33}, 37 (2003), arXiv:gr-qc/0210013.

\bibitem{UhlenbeckOrnstein1930}

G. E. Uhlenbeck and L. S. Ornstein, Phys. Rev. \textbf{36}, 823 (1930).

\end{thebibliography}
\end{document}